\input harvmac
\input amssym

\def\Dslash{D\kern -0.67em/}

\def\zb{\overline{z}}

 \def\mb{\bar{m}}

\def\wb{{\overline{w}}}

\def\wh{{\hat{w}}}

\baselineskip 13pt

\lref\LarsenXM{
  F.~Larsen,
  ``The Perturbation spectrum of black holes in N=8 supergravity,''
Nucl.\ Phys.\ B {\bf 536}, 258 (1998).
[hep-th/9805208].
}
\lref\MichelsonKN{
  J.~Michelson and M.~Spradlin,
  ``Supergravity spectrum on AdS(2) x S**2,''
JHEP {\bf 9909}, 029 (1999).
[hep-th/9906056].
}
\lref\LeeYU{
  J.~Lee and S.~Lee,
  ``Mass spectrum of D = 11 supergravity on AdS(2) x S**2 x T**7,''
Nucl.\ Phys.\ B {\bf 563}, 125 (1999).
[hep-th/9906105].
}
\lref\CorleyUZ{
  S.~Corley,
  ``Mass spectrum of N=8 supergravity on AdS(2) x S**2,''
JHEP {\bf 9909}, 001 (1999).
[hep-th/9906102].
}

\lref\SalamXD{
  A.~Salam and J.~A.~Strathdee,
  ``On Kaluza-Klein Theory,''
Annals Phys.\  {\bf 141}, 316 (1982).
}

\lref\GrumillerNM{
  D.~Grumiller, W.~Kummer and D.~V.~Vassilevich,
Phys.\ Rept.\  {\bf 369}, 327 (2002).
[hep-th/0204253].
}

\lref\AbrikosovNJ{
  A.~A.~Abrikosov, Jr.,
  ``Fermion states on the sphere S**2,''
Int.\ J.\ Mod.\ Phys.\ A {\bf 17}, 885 (2002).
[hep-th/0111084].
}

\lref\VassilevichXT{
  D.~V.~Vassilevich,
  ``Heat kernel expansion: User's manual,''
Phys.\ Rept.\  {\bf 388}, 279 (2003).
[hep-th/0306138].
}

\lref\CamporesiWN{
  R.~Camporesi and A.~Higuchi,
  ``Stress energy tensors in anti-de Sitter space-time,''
Phys.\ Rev.\ D {\bf 45}, 3591 (1992).
}

\lref\CamporesiGA{
  R.~Camporesi and A.~Higuchi,
  ``Spectral functions and zeta functions in hyperbolic spaces,''
J.\ Math.\ Phys.\  {\bf 35}, 4217 (1994).
}

\lref\CamporesiFB{
  R.~Camporesi and A.~Higuchi,
J.\ Geom.\ Phys.\  {\bf 20}, 1 (1996).
[gr-qc/9505009].
}

\lref\BanerjeeQC{
  S.~Banerjee, R.~K.~Gupta and A.~Sen,
  ``Logarithmic Corrections to Extremal Black Hole Entropy from Quantum Entropy Function,''
JHEP {\bf 1103}, 147 (2011).
[arXiv:1005.3044 [hep-th]].
}

\lref\BanerjeeJP{
  S.~Banerjee, R.~K.~Gupta, I.~Mandal and A.~Sen,
  ``Logarithmic Corrections to N=4 and N=8 Black Hole Entropy: A One Loop Test of Quantum Gravity,''
JHEP {\bf 1111}, 143 (2011).
[arXiv:1106.0080 [hep-th]].
}

\lref\SenBA{
  A.~Sen,
  ``Logarithmic Corrections to N=2 Black Hole Entropy: An Infrared Window into the Microstates,''
[arXiv:1108.3842 [hep-th]].
}

\lref\SenAJA{
  A.~Sen,
  ``Microscopic and Macroscopic Entropy of Extremal Black Holes in String Theory,''
[arXiv:1402.0109 [hep-th]].
}

\lref\NicolaiTD{
  H.~Nicolai and P.~K.~Townsend,
  ``N=3 Supersymmetry Multiplets with Vanishing Trace Anomaly: Building Blocks of the $N\ge3$ Supergravities,''
Phys.\ Lett.\ B {\bf 98}, 257 (1981).
}

\lref\FerraraIH{
  S.~Ferrara, R.~Kallosh and A.~Strominger,
  ``N=2 extremal black holes,''
Phys.\ Rev.\ D {\bf 52}, 5412 (1995).
[hep-th/9508072].
}

\lref\HawkingJA{
  S.~W.~Hawking,
  ``Zeta Function Regularization of Path Integrals in Curved Space-Time,''
Commun.\ Math.\ Phys.\  {\bf 55}, 133 (1977)..
}

\lref\ChristensenMD{
  S.~M.~Christensen and M.~J.~Duff,
  ``New Gravitational Index Theorems and Supertheorems,''
Nucl.\ Phys.\ B {\bf 154}, 301 (1979)..
}

\lref\GuptaHXA{
  R.~K.~Gupta, S.~Lal and S.~Thakur,
  ``Logarithmic Corrections to Extremal Black Hole Entropy in N = 2, 4 and 8 Supergravity,''
[arXiv:1402.2441 [hep-th]].
}

\lref\GuptaSVA{
  R.~K.~Gupta, S.~Lal and S.~Thakur,
  ``Heat Kernels on the $AdS_2$ cone and Logarithmic Corrections to Extremal Black Hole Entropy,''
JHEP {\bf 1403}, 043 (2014).
[arXiv:1311.6286 [hep-th]].
}

\lref\ArosTAA{
  R.~Aros, D.~E.~Diaz and A.~Montecinos,
  ``On Wald entropy of black holes: logarithmic corrections and trace anomaly,''
[arXiv:1305.4647 [gr-qc]].
}

\lref\PourdarvishGFA{
  A.~Pourdarvish, J.~Sadeghi, H.~Farahani and B.~Pourhassan,
  ``Thermodynamics and Statistics of Goedel Black Hole with Logarithmic Correction,''
Int.\ J.\ Theor.\ Phys.\  {\bf 52}, 3560 (2013).
}

\lref\BhattacharyyaWZ{
  S.~Bhattacharyya, B.~Panda and A.~Sen,
  ``Heat Kernel Expansion and Extremal Kerr-Newmann Black Hole Entropy in Einstein-Maxwell Theory,''
JHEP {\bf 1208}, 084 (2012).
[arXiv:1204.4061 [hep-th]].
}

\lref\AbrikosovJR{
  A.~A.~Abrikosov, Jr.,
  ``Dirac operator on the Riemann sphere,''
[hep-th/0212134].
}

\lref\DasIC{
  S.~Das, P.~Majumdar and R.~K.~Bhaduri,
  ``General logarithmic corrections to black hole entropy,''
Class.\ Quant.\ Grav.\  {\bf 19}, 2355 (2002).
[hep-th/0111001].
}

\lref\SenDW{
  A.~Sen,
  ``Logarithmic Corrections to Schwarzschild and Other Non-extremal Black Hole Entropy in Different Dimensions,''
JHEP {\bf 1304}, 156 (2013).
[arXiv:1205.0971 [hep-th]].
}

\lref\SenCJ{
  A.~Sen,
  ``Logarithmic Corrections to Rotating Extremal Black Hole Entropy in Four and Five Dimensions,''
Gen.\ Rel.\ Grav.\  {\bf 44}, 1947 (2012).
[arXiv:1109.3706 [hep-th]].
}

\lref\WaldNT{
  R.~M.~Wald,
  ``Black hole entropy is the Noether charge,''
Phys.\ Rev.\ D {\bf 48}, 3427 (1993).
[gr-qc/9307038].
}

\lref\JacobsonVJ{
  T.~Jacobson, G.~Kang and R.~C.~Myers,
  ``On black hole entropy,''
Phys.\ Rev.\ D {\bf 49}, 6587 (1994).
[gr-qc/9312023].
}

\lref\CamporesiNW{
  R.~Camporesi,
  ``zeta function regularization of one loop effective potentials in anti-de Sitter space-time,''
Phys.\ Rev.\ D {\bf 43}, 3958 (1991)..
}

\lref\SenVM{
  A.~Sen,
  ``Quantum Entropy Function from AdS(2)/CFT(1) Correspondence,''
Int.\ J.\ Mod.\ Phys.\ A {\bf 24}, 4225 (2009).
[arXiv:0809.3304 [hep-th]].
}

\lref\StromingerKF{
  A.~Strominger,
  ``Macroscopic entropy of N=2 extremal black holes,''
Phys.\ Lett.\ B {\bf 383}, 39 (1996).
[hep-th/9602111].
}

\lref\GopakumarQS{
  R.~Gopakumar, R.~K.~Gupta and S.~Lal,
  ``The Heat Kernel on $AdS$,''
JHEP {\bf 1111}, 010 (2011).
[arXiv:1103.3627 [hep-th]].
}

\lref\KeelerBRA{
  C.~Keeler, F.~Larsen and P.~Lisbao,
  ``Logarithmic Corrections to $N \geq 2$ Black Hole Entropy,''
[arXiv:1404.1379 [hep-th]].
}

\lref\DavidXG{
  J.~R.~David, M.~RGaberdiel and R.~Gopakumar,
  ``The Heat Kernel on AdS(3) and its Applications,''
JHEP {\bf 1004}, 125 (2010).
[arXiv:0911.5085 [hep-th]].
}

\lref\GiombiVD{
  S.~Giombi, A.~Maloney and X.~Yin,
  ``One-loop Partition Functions of 3D Gravity,''
JHEP {\bf 0808}, 007 (2008).
[arXiv:0804.1773 [hep-th]].
}

\lref\BhattacharyyaYE{
  S.~Bhattacharyya, A.~Grassi, M.~Marino and A.~Sen,
  ``A One-Loop Test of Quantum Supergravity,''
Class.\ Quant.\ Grav.\  {\bf 31}, 015012 (2014).
[arXiv:1210.6057 [hep-th]].
}

\lref\deBoerVF{
  J.~de Boer and S.~N.~Solodukhin,
  ``A Holographic reduction of Minkowski space-time,''
Nucl.\ Phys.\ B {\bf 665}, 545 (2003).
[hep-th/0303006].
}

\lref\HeCRA{
  T.~He, P.~Mitra, A.~P.~Porfyriadis and A.~Strominger,
  ``New Symmetries of Massless QED,''
[arXiv:1407.3789 [hep-th]].
}

\lref\HeLAA{
  T.~He, V.~Lysov, P.~Mitra and A.~Strominger,
  ``BMS supertranslations and Weinberg's soft graviton theorem,''
[arXiv:1401.7026 [hep-th]].
}

\lref\StromingerJFA{
  A.~Strominger,
  ``On BMS Invariance of Gravitational Scattering,''
[arXiv:1312.2229 [hep-th]].
}

\lref\ElitzurBF{
  S.~Elitzur, A.~Forge and E.~Rabinovici,
  ``Comments on the importance of being overconstrained,''
Phys.\ Lett.\ B {\bf 289}, 45 (1992)..
}

\lref\BuchbinderNIA{
  E.~I.~Buchbinder and A.~A.~Tseytlin,
  ``The 1/N correction in the D3-brane description of circular Wilson loop at strong coupling,''
Phys.\ Rev.\ D {\bf 89}, 126008 (2014).
[arXiv:1404.4952 [hep-th]].
}

\lref\FaraggiTNA{
  A.~Faraggi, J.~T.~Liu, L.~A.~Pando Zayas and G.~Zhang,
  ``One-loop Agreement of Higher Rank Wilson Loops in AdS/CFT,''
[arXiv:1409.3187 [hep-th]].
}
\lref\KruczenskiBLA{
  M.~Kruczenski,
  ``Wilson loops and minimal area surfaces in hyperbolic space,''
[arXiv:1406.4945 [hep-th]].
}

\lref\BrittoPacumioAX{
  R.~Britto-Pacumio, J.~Michelson, A.~Strominger and A.~Volovich,
  ``Lectures on Superconformal Quantum Mechanics and Multi-Black Hole Moduli Spaces,''
NATO Sci.\ Ser.\ C {\bf 556}, 255 (2000).
[hep-th/9911066].
}

\lref\CastroIMA{
  A.~Castro and W.~Song,
  ``Comments on AdS$_2$ Gravity,''
[arXiv:1411.1948 [hep-th]].
}

\lref\SenVM{
  A.~Sen,
  ``Quantum Entropy Function from AdS(2)/CFT(1) Correspondence,''
Int.\ J.\ Mod.\ Phys.\ A {\bf 24}, 4225 (2009).
[arXiv:0809.3304 [hep-th]].
}

\lref\CastroMS{
  A.~Castro, D.~Grumiller, F.~Larsen and R.~McNees,
  ``Holographic Description of AdS(2) Black Holes,''
JHEP {\bf 0811}, 052 (2008).
[arXiv:0809.4264 [hep-th]].
}

\lref\BanerjeeQC{
  S.~Banerjee, R.~K.~Gupta and A.~Sen,
  ``Logarithmic Corrections to Extremal Black Hole Entropy from Quantum Entropy Function,''
JHEP {\bf 1103}, 147 (2011).
[arXiv:1005.3044 [hep-th]].
}

\lref\ChristensenIY{
  S.~M.~Christensen and M.~J.~Duff,
Nucl.\ Phys.\ B {\bf 170}, 480 (1980)..
}

\lref\GibbonsJI{
  G.~W.~Gibbons and M.~J.~Perry,
Nucl.\ Phys.\ B {\bf 146}, 90 (1978)..
}

\lref\FaraggiBB{
  A.~Faraggi and L.~A.~Pando Zayas,
JHEP {\bf 1105}, 018 (2011).
[arXiv:1101.5145 [hep-th]].
}

\lref\KruczenskiZK{
  M.~Kruczenski and A.~Tirziu,
JHEP {\bf 0805}, 064 (2008).
[arXiv:0803.0315 [hep-th]].
}

\Title{\vbox{\baselineskip12pt\
}}
{\vbox{\centerline{Quantum Corrections to Supergravity on AdS$_2\times S^2$}}}
\centerline{
Finn
Larsen\foot{larsenf@umich.edu}, and Pedro Lisb\~{a}o\foot{plisbao@umich.edu}
}
\bigskip
\centerline{\it{Department of Physics
and Michigan Center for Theoretical Physics,}}
\centerline{\it{University of Michigan, Ann Arbor, MI 48109-1120, USA.}}

\baselineskip15pt

\vskip .3in

\centerline{\bf Abstract}
We compute the off-shell spectrum of supergravity on AdS$_2\times S^2$ by explicit diagonalization of the equations of motion for an effective AdS$_2$ theory where all fields are dualized to scalars and spin-${1\over 2}$ fermions. Classifying all bulk modes as physical, gauge violating, and pure gauge let us identify boundary modes 
as physical fields on $S^2$ that are formally pure gauge but with gauge function that is non-normalizable on AdS$_2$. As an application we compute the leading quantum correction to AdS$_2\times S^2$ as a sum over physical fields including boundary states. 

\Date{November, 2014}
\baselineskip14pt

\newsec{Introduction}
Quantum corrections to solutions of general relativity are computed by Gaussian integrals over the quadratic fluctuations 
around the gravitational background. Regularization and renormalization of the resulting functional determinants were carried 
out explicitly a long time ago for many general settings using heat kernel methods, zeta-function techniques and others. 
However, modern applications of the AdS/CFT correspondence usually embed solutions into supergravity and these settings typically activate many fields with non-minimal couplings. This situation presents new conceptual challenges and it also focusses attention on 
unresolved difficulties in the literature. 

Supergravity couplings organize physical states efficiently according to quantum numbers such as conformal dimension. However, 
unphysical modes are often unwieldy since auxiliary fields and ghosts involved in the off-shell theory also couple non-minimally. 
These complications seem excessive for determinants of quadratic fluctuations so it may be advantageous to work in the small Hilbert 
space that focusses entirely on the physical modes. The resulting on-shell strategy is simpler but it must address global aspects that remain after gauge fixing of local symmetries. Specifically, there will be boundary modes in AdS. 

In this paper we develop the on-shell method in the context of supergravity on  AdS$_2 \times S^2$.  Our results for quantum corrections are not new as they were previously reported in \refs{\SenBA,\GuptaHXA,\KeelerBRA} but we  present explicit details that develop concepts and resolve issues in the literature. 

An important motivation for developing quantum corrections in AdS$_2$ and specifically the role of boundary modes is that they play a central role also in other settings. Some recent discussions are:
\itemitem{$\bullet$}
Boundary states are standard in AdS$_3$ partition functions \refs{\GiombiVD,\DavidXG} and they presumably play a similar role in higher dimensional AdS spaces \refs{\GopakumarQS,\BhattacharyyaYE}. 
\itemitem{$\bullet$}
Quantum corrections in AdS$_2$ geometry appear for Wilson loops in AdS$_5$ \BuchbinderNIA. Subtleties remain in this context \refs{\FaraggiTNA, \FaraggiBB, \KruczenskiBLA, \KruczenskiZK}. 
\itemitem{$\bullet$}
AdS$_2\times S^2$ is conformally equivalent to Minkowski space so these modes may also be related to the physical boundary modes that play a role in scattering amplitudes \refs{\StromingerJFA,\HeLAA,\HeCRA} and to those that appear in the context of holography in Minkowski space \refs{\deBoerVF}. 
\itemitem{$\bullet$}
Our set-up is an explicit realization of AdS$_2/$CFT$_1$ holography. Many open questions remain in this context \refs{\BrittoPacumioAX,\CastroMS,\SenVM,\CastroIMA}.

In our computation we organize the field content on AdS$_2$ into towers of partial waves due to the reduction on the $S^2$. We analyze this 2D spectrum with gauge fixing terms included in the equations of motion but not imposed as constraints. In our presentation we explicitly 
identify some towers as unphysical (they violate the gauge condition) and others as pure gauge (the action of ${\rm diff} \times {\rm gauge}$ on the background), with the remaining fields constituting the physical bulk spectrum. Equivalenly, we match both the unphysical and gauge towers with ghosts and determine the ``small'' departure from perfect cancellation. In either construction, the bulk spectrum is thus augmented by physical modes that are formally pure gauge albeit with non-normalizable gauge function. These are the boundary modes.

In our construction each local symmetry in 4D gives rise to a tower of boundary modes in AdS$_2$. We interpret such a tower as a single field on $S^2$. There is exactly one such boundary field on $S^2$ for each symmetry. It may appear that we have lost a dimension: the boundary of AdS$_2\times S^2$ has one dimension, in addition to the $S^2$ dimensions. Indeed, at an intermediate stage there is one mode for each boundary momentum on AdS$_2$ but we reinterpret the resulting sum as the volume of 
AdS$_2$. It is in this sense that we find exactly one mode on $S^2$ for each 4D symmetry.

We express quantum corrections to the geometry as heat kernel sums over the spectrum. In the ``large" Hilbert space these are traces over the full spectrum with unphysical modes cancelled by ghosts with ``wrong" statistics. These sums can be reorganized as traces over the physical spectrum in the ``small" Hilbert space where boundary states are included and all modes appear with a positive sign. The boundary fields include components that are zero-modes on AdS$_2\times S^2$ and such modes require special treatment \BanerjeeQC. The complete partition function thus comprises modes in 4D (bulk), 2D (boundary), and 0D (zero-modes).  

The main idea of our computation can be illustrated clearly by considering a standard (minimally coupled) vector field $A_I$ in AdS$_2\times S^2$. The partial wave expansion on $S^2$ gives four towers of 2D fields: two physical (spatially transverse), one unphysical (violating the gauge condition), and one longitudinal (pure gauge). In the old-fashioned Gupta-Bleuler formalism the unphysical and the longitudinal towers ``cancel" (due to a Ward identity) and in BRST formalism both towers are cancelled by ghosts. Either way, for each partial wave the mode that is formally pure gauge can be arranged to require a non-normalizable gauge function on AdS$_2$ and this gives rise to a single physical longitudinal mode that survives as an AdS$_2$ boundary mode. 

Standard AdS/CFT lore sometimes suggests that physical boundary states are at the ``end" of the physical towers but we find this rule to be misleading. Indeed, since boundary states arise formally as states that are pure gauge it may be more appropriate to interpret them as the ``end" of the {\it unphysical} towers. However, ultimately it turns out that couplings between boundary modes render such shortcuts unreliable. One aspect of this is that modes generated by symmetries generally do not continue smoothly from general partial wave component $l$ to the ``small" values $l=0,1$. 

As we have indicated, boundary states can be interpreted as modes that are formally  ``pure gauge". An alternative perspective ties them to harmonic modes on AdS$_2$ which play a special role when fields of higher spin are dualized to scalars. We find that the dual of gravity includes an interesting harmonic scalar satisfying a higher order equation of motion with solutions for both $m^2=0$ and $m^2=2$. It is the latter that gives rise to physical boundary modes for gravity. 
This twist on the harmonic condition may be significant in other settings. 

The detailed considerations are instructive but they are unfortunately somewhat cumbersome even in the simple example of AdS$_2\times S^2$. That is a byproduct of analyzing ${\cal N}=2$ supergravity off-shell without introducing a full-fledged off-shell formalism. Several asymmetries give rise to a non-Hermitian action for off-shell fields which manifests itself by awkward degenerate eigenvectors. For example, diffeomorphisms act on gauge fields but gauge transformations do not act on the metric. The pay-off for addressing these practical complications is considerable conceptual clarity. 

This article is organized as follows. 
In section 2 we present the details of a minimally coupled vector field on AdS$_2\times S^2$. We reduce from 4D to 2D, diagonalize the off-shell 2D equations in Lorentz gauge, and discuss the physical spectrum. We specify the boundary modes as pure gauge modes with non-normalizable gauge function and also as harmonic modes. 
In section 3 we compute the heat kernel of the vector field as a sum over all physical states in bulk and on the boundary. We compare with the standard off-shell computation. In section 4 we discuss the analogous aspects of the bosonic fields in the ${\cal N}=2$ supergravity multiplet. We also address additional features: degenerate eigenvalues and modes, the harmonic condition on the scalar dual to a tensor field, residual 2D diffeomorphism invariance, and the role of (Conformal) Killing Vectors. In section 5, we discuss the heat kernels of the bosonic fields with special emphasis on the cancellation of off-shell modes and the contribution of physical boundary states. 
In section 6 we turn to the gravitinos in the ${\cal N}=2$ supergravity multiplet. We again diagonalize the equations of motion entirely without any gauge condition imposed and only then discuss supersymmetry and the constraints inherent in the Rarita-Schwinger equation. Finally, in section 7 we compute the heat kernel for the gravitini an assemble the full result for supergravity on AdS$_2\times S^2$. 

\newsec{Vector Fields in AdS$_2\times S^2$}
In this section we analyze a vector field in AdS$_2\times S^2$ from the AdS$_2$ point of view. We determine the full set of modes in 4D Lorentz gauge and identify the physical subset with special attention paid to the boundary modes. 

\subsec{The 2D Effective Theory}


Our starting point is a 4D vector field $a_I$ on AdS$_2\times S^2$ with standard Maxwell action
\eqn\aa{
{\cal L}_{\rm Maxwell} = -{1\over 4} F_{IJ} F^{IJ}~. 
}
In order to extract the physical content of the theory we impose Lorentz gauge
\eqn\ab{
\nabla_I a^I =0~.
}
In the quantum theory this is implemented by modifying the Maxwell action \aa\ to 
\eqn\ac{
{\cal L}_{\rm Lorentz} = -{1\over 4} F_{IJ} F^{IJ} - {1\over 2\xi} \left( \nabla_I a^I\right)^2  ~. 
}
In the following we take Feynman gauge $\xi=1$ and freely integrate by parts without keeping boundary terms. The action then simplifies to
\eqn\ad{\eqalign{
{\cal L}_{\rm Feynman} = {1\over 2} a^J \nabla^I (\nabla_I a_J - \nabla_J a_I )  +  {1\over 2} a^J \nabla_J \nabla_I a^I  
= {1\over 2} a^I (g_{IJ} \nabla^2  - R_{IJ} )  a^J~. 
}}

We want to represent this theory as an effective theory in 2D by reduction on $S^2$. In so doing the capital latin indices $I,J,\ldots$ in the 4D total space divide into the indices $\mu,\nu,\ldots$ on AdS$_2$ and the indices $\alpha,\beta,\ldots$ that refer to $S^2$. 
The reduction to 2D on $S^2$ is realized by a partial wave expansion in spherical harmonics:
\eqn\ae{\eqalign{
a_\mu & = b^{(lm)}_\mu (x) Y_{lm}(y)~,\cr
a_\alpha & = b^{(lm)} (x) \epsilon_{\alpha\beta}\nabla^\beta Y_{lm}(y)
+\tilde{b}^{(lm)} (x) \nabla_\alpha Y_{lm}(y)~.
}}
 A sum over angular momentum quantum numbers $l,m$ is implied. The allowed angular momenta for the 2D gauge fields $b^{(lm)}_\mu$ are $l=0,1,\ldots$ but the 2D scalar fields 
$b^{(lm)} (x), \tilde{b}^{(lm)} (x)$ are defined only for $l=1,2,\ldots$ since these fields multiply spherical harmonics with derivatives acting on them. 

Inserting the expansions \ae\ into the 4D Lagrangian \ad\ we find the 2D effective action on AdS$_2$ 
\eqn\af{\eqalign{
{\cal L}_{2D} = & {1\over 2} l(l+1)b^{(lm)}\left[ \nabla^2_A -l(l+1)\right]b^{(lm)}   + {1\over 2} l(l+1)\tilde{b}^{(lm)} 
\left[ \nabla^2_A -l(l+1)\right] \tilde{b}^{(lm)} \cr
&+ {1\over 2} b^{(lm)\mu} \left[ \nabla^2_A + 1 - l(l+1)  \right] b^{(lm)}_\mu~.
}}
The 2D Laplacian on AdS$_2$ is denoted $\nabla^2_A = \nabla^\mu\nabla_\mu$. We still imply a sum over fields $l=0,1,\ldots$. This rule correctly takes into account that the $l=0$ mode is missing for $b^{(lm)}$ and $\tilde{b}^{(lm)}$  but it is not missing for $b_\mu^{(lm)}$. Curvature terms from commutation of derivatives were evaluate using the block diagonal Ricci tensor with $R_{\mu\nu} = -g_{\mu\nu}$ and $R_{\alpha\beta} = +g_{\alpha\beta}$ of AdS$_2\times S^2$ with unit radii.

The gauge variation of the Lorentz gauge condition \ab\ is
\eqn\bq{
\nabla^I\delta A_I = \nabla^I \nabla_I \Lambda = (\nabla^2_A - l(l+1))\Lambda~.
}
We will variously interpret this as the equation of motion for the pure gauge mode or as the
ghost action
\eqn\br{
{\cal L}_{\rm ghost} = \tilde{c}^{(lm)}(\nabla^2_A - l(l+1)) c^{(lm)}~.
}
The ghost spectrum $m^2=l(l+1)$ with $l=0,1,\ldots$ is identical to two scalar fields except for anti-commuting statistics. 

\subsec{Dualizing 2D Vectors}
The Hodge decomposition of a 1-form into an exact form, a co-exact form, and a harmonic form can be presented in components as
\eqn\ba{\eqalign{
b_\mu^{(lm)} & = b_{\mu\perp}^{(lm)} + b_{\mu\parallel}^{(lm)} + b^{(lm)}_{\mu0}~,
}}
where $b_{\mu\perp}^{(lm)}$ is transverse
\eqn\bb{
\nabla^\mu b_{\mu\perp}^{(lm)} =0~,
}
and $b_{\mu\parallel}^{(lm)}$ is longitudinal
\eqn\bc{
\epsilon^{\mu\nu}\nabla_\nu b_{\mu\parallel}^{(lm)} =0~,
}
while $b^{(lm)}_{\mu0}$ satisfies both of the above. In order to avoid over counting of modes we insist that 
\eqn\bd{
\nabla^\mu b_{\mu\parallel}^{(lm)} \neq 0~~~~,~
\epsilon^{\mu\nu}\nabla_\nu b_{\mu\perp}^{(lm)} \neq 0~. 
}
This is because the modes satisfying both of \bb\ and \bc\ are the harmonic modes denoted $b_{\mu 0}^{(lm)}$. The harmonic component of the vector field satisfies 
\eqn\bia{
(\nabla^2_A+1) b^{(lm)}_{\mu 0} =0~.
}

We dualize the irreducible components of the 2D vector $b_{\mu}^{(lm)}$ to scalars as $b_{\mu\perp}^{(lm)} = \epsilon_{\mu\nu}\nabla^\nu b_{\perp}^{(lm)}$ and $b_{\mu\parallel}^{(lm)}=\nabla_\mu b_{\parallel}^{(lm)}$. This gives
the expansion 
\eqn\be{\eqalign{
b_\mu^{(lm)}  &= \epsilon_{\mu\nu}\nabla^\nu b_{\perp}^{(lm)}+ \nabla_\mu b_{\parallel}^{(lm)} +  \nabla_\mu b_{0}^{(lm)}~, 
}}
For definiteness the harmonic mode was presented as a longitudinal mode $b^{(lm)}_{\mu0}=\nabla_\mu b^{(lm)}_0$ with $b^{(lm)}_0$ harmonic
\eqn\bi{
\nabla^2_A b^{(lm)}_0 =0~,
}
but we might as well have dualized it to a transverse mode. In our convention the scalar components $b_{\parallel}^{(lm)}$ and $b_{\perp}^{(lm)}$ cannot be harmonic on AdS$_2$.

\subsec{The Spectrum}
The complete field content of the 4D vector field from a 2D point of view is:

\medskip

\itemitem{$\bullet$}
{\bf Modes on $S^2$:}
$\tilde{b}^{(lm)} , b^{(lm)}$ with $l=1,2,\ldots$

\itemitem{$\bullet$}
{\bf Modes on AdS$_2$:}
$b_{\mu\perp}^{(lm)} = \epsilon_{\mu\nu}\nabla^\nu b_{\perp}^{(lm)}$ and $b_{\mu\parallel}^{(lm)}=\nabla_\mu b_{\parallel}^{(lm)}$ with $l=0,1,\ldots$ 

\itemitem{$\bullet$}
{\bf Ghosts:}
$\tilde{c}^{(lm)}, c^{(lm)}$ with $l=0,1,\ldots$ 

\itemitem{$\bullet$}
{\bf Harmonic modes:}
$b^{(lm)}_{\mu 0}=\nabla_\mu b^{(lm)}_0$ with $l=0,1,\ldots$

In the fully dualized theory there is almost symmetry between AdS$_2$ and $S^2$ after appropriate interpretations. One departure from perfect symmetry is the ``subtraction'' of the leading $l=0$ entry from the scalars $b^{(lm)}, \tilde{b}^{(lm)}$ which represent the vector on $S^2$ that only has range $l=1,2,\ldots $. This contrasts with the scalars $b_{\parallel}^{(lm)},b_{\perp}^{(lm)}$ from the AdS$_2$ vector. These have the full range $l=0,1,\ldots $ and also ``add" the harmonic fields $b^{(lm)}_0$. 

Each 2D field is a scalar field on AdS$_2$ with mass given by $m^2=l(l+1)$. At the level of counting, the modes on AdS$_2$ cancel exactly with the ghosts. The net physical spectrum is therefore essentially just the modes on $S^2$ forming two towers with $l=1,2,\ldots$. These correspond to the partial wave expansions of two physical modes with helicity $\lambda=\pm 1$ that we expect from a 4D vector field. 

It is instructive to go beyond counting and construct physical modes explicitly. We first assume $l\geq 1$ and consider the gauge condition $\ab$. It amounts to 
\eqn\ce{
\nabla^\mu b^{(lm)}_{\mu\parallel} - l(l+1) \tilde{b}^{(lm)}=
\nabla^2_A b^{(lm)}_{\parallel} - l(l+1) \tilde{b}^{(lm)}=0~,
}
in terms of 2D modes. Only one linear combination of the modes $b^{(lm)}_{\parallel}$, $\tilde{b}^{(lm)}$ satisfies the gauge condition. On-shell the equations of motion impose $\nabla^2_A b^{(lm)}_{\parallel} = l(l+1)b^{(lm)}_{\parallel}$ so the physical modes are those that satisfy $\tilde{b}^{(lm)}=b^{(lm)}_{\parallel}$. 

We next consider the 4D gauge symmetry $a_I\to a_I + \nabla_I \Lambda$. Expanding the gauge function $\Lambda$ in spherical 
harmonics
\eqn\cj{
\Lambda = \lambda^{(lm)}(x) Y_{lm}(y)~,
}
this amounts to the 2D transformations
\eqn\ck{\eqalign{
\tilde{b}^{(lm)} & \to \tilde{b}^{(lm)} + \lambda^{(lm)}~,
\cr
b^{(lm)}_{\mu\parallel} & \to b^{(lm)}_{\mu\parallel} + \nabla_\mu \lambda^{(lm)}~.
}}
The field configurations identified after \ce\ as satisfying the gauge condition on-shell have $\tilde{b}^{(lm)}=b^{(lm)}_{\parallel}$ with $b_{\mu\parallel}^{(lm)}=\nabla_\mu b^{(lm)}_{\parallel}$. Therefore these are precisely those that are gauge equivalent to the vacuum. Such pure gauge configurations decouple from processes involving states that do satisfy the gauge condition. 

The modes $b^{(lm)}$ and $b^{(lm)}_{\mu\perp}=\epsilon_{\mu\nu} \nabla^\nu b^{(lm)}_{\perp}$ do not enter the gauge conditions \ce\ at all, nor are they acted on by the gauge transformations \ck. These therefore form two towers of physical modes. Since we assumed $l\geq 1$ from the outset the range of these towers is $l=1,2,\cdots$ as expected. 

The lowest spherical harmonic $l=0$ requires special consideration. Indeed, the scalar fields $b^{(00)}, \tilde{b}^{(00)}$ from the $S^2$ components of the vector field are non-existent because partial waves on $S^2$ have $l\geq 1$. Further, for $l=0$ the on-shell condition on the scalars $b^{(00)}_{\parallel},b^{(00)}_{\perp}$ due to the AdS$_2$ components of the vector field reduces to the harmonic condition on AdS$_2$ and in \bd\ we specifically exempt harmonic modes. Thus there are no modes at $l=0$ before even considering the gauge condition and the possibility of pure gauge modes. 

In summary, the more detailed discussion identifies the physical modes as the towers $b^{(lm)}$, $b^{(lm)}_{\perp}$ with $l=1,2,\ldots$. Importantly, these are not simply the modes $b^{(lm)}$, $\tilde{b}^{(lm)}$ that were defined with range $l=1,2,\ldots$ from the outset. Indeed, the mode $b^{(lm)}_{\perp}$ was defined for $l=0,1,\ldots$ but the harmonic condition removed the $l=0$ entry.

\subsec{Boundary Modes}
The discussion of the spectrum so far deferred consideration of the harmonic modes $b^{(lm)}_0$ introduced in \be. These give rise to boundary modes. Several comments are in order: 

\itemitem{$\bullet$}
There is exactly one harmonic mode for each partial wave $(lm)$: the AdS$_2$  vector $b^{(lm)}_\mu$ is dualized to two scalar  components  $b^{(lm)}_{\perp}$ and $b^{(lm)}_{\parallel}$ but the harmonic mode $b^{(lm)}_0$ is ``shared'' between these fields as it is both longitudinal and transverse. 

\itemitem{$\bullet$}
The tower of harmonic modes begins at $l=0$ just like all other components of the AdS$_2$  vector.

\itemitem{$\bullet$}
The harmonic condition implies that these modes are zero-modes on AdS$_2$. The tower of harmonic modes --- one for each $(lm)$ --- identifies the configuration space of harmonic modes as a field on $S^2$. The equation of motion of this field identifies the leading $l=0$ mode as physical. 

\itemitem{$\bullet$}
The scalar Laplacian $(-\nabla^2_A)$ in Euclidean AdS$_2$ has eigenvalues $c_2 = {1\over 4} + s^2$ with $s$ real for fields in the principal continuous representations of $SL(2)$. These representations are AdS$_2$ analogues of plane waves in flat space. The harmonic mode has $c_2=0$ and belongs to a principal discrete representation with no flat space analogue.

\itemitem{$\bullet$}
The harmonic modes are formally pure gauge since they are longitudinal. However, they are physical because the gauge function that generates them is non-normalizable. For us the term harmonic mode is synonymous with the term boundary mode because AdS/CFT lore posits that pure gauge degrees of freedom localize on the boundary. 

The harmonic modes were constructed explicitly some time ago \CamporesiGA. In our discussion we write the Euclidean AdS$_2$ black hole metric in complex form as
\eqn\bt{
ds^2_2 = a^2 ( d\eta^2 + \sinh^2\eta d\theta^2 )  = a^2 {4\over (1-|z|^2)^2} dz d\bar z ~,
}
where $\theta$ has period $2\pi$ and $z=\tanh {\eta\over 2}e^{i\theta}$. The conformal factor in the second expression diverges as the AdS$_2$ boundary $|z| = 1$ is approached but this does not affect the harmonic condition which is conformally invariant. We can therefore choose a standard complete set of harmonic modes such as\foot{We omit the constant on AdS$_2$ (corresponding to $n=0$) since only derivatives of the basis parametrize vector fields.}
\eqn\bu{
u_n = {1\over \sqrt{2\pi n}} z^n ~,~~~~n=1,2\ldots~,
}
and their complex conjugates. These modes cannot appear as components of a scalar field on AdS$_2$ since the normalization condition
\eqn\bw{
\int \sqrt{g} d^2 z ~ |u_n |^2 = \int  {2a^2 d^2 z\over (1-|z|^2)^2} ~ |u_n |^2  \to \infty~,
}
diverges at the boundary due to the conformal factor. However, derivatives of the modes \bu\ are subject to a conformally invariant normalization condition so they are legitimate components of a vector field. The modes \bu\ are normalized so
\eqn\bv{
\int  \sqrt{g} d^2 z ~|\nabla_z u_n |^2= 1~,
}
in standard conventions where $d^2 z= 2dxdy$. Vector fields formed from gradients of harmonic modes are therefore physical even though they are formally pure gauge. We interpret them as boundary modes.

\subsec{BRST Quantization}
Our old-fashioned discussion of physical modes extends immediately to the more streamlined BRST quantization. For completeness we briefly outline this generalization. 

The physical fields $b^{(lm)}_{\perp}, b^{(lm)}$ are BRST invariant. Other BRST invariant field configurations are those that have no anti-ghosts $\tilde{c}^{(lm)}=0$ and also satisfy $\tilde{b}^{(lm)}=b^{(lm)}_{\parallel}$. 

The ghost states $c^{(lm)}$ are BRST exact since they are BRST transforms of pure gauge fields. The gauge fields with $\tilde{b}^{(lm)}=b^{(lm)}_{\parallel}$ are also BRST exact since they are BRST transforms of anti-ghosts $\tilde{c}^{(lm)}$. 

This accounting leaves just the physical fields $b^{(lm)}_{\perp}, b^{(lm)}$ with $l=1,2,\ldots$. 

The spherically symmetric fields $l=0$ must be considered separately. The antighost fails to be BRST invariant and the ghost is the BRST transform of a pure gauge function. The remaining two fields $b^{(00)}_{\perp}, b^{(00)}_{\parallel}$ are not independent on-shell and can be formally presented as the BRST transform of the anti-ghost $\tilde{c}^{(00)}$, albeit with a non-normalizable field configuration. 

In summary, the BRST cohomology agrees with the physical states discussed above in a more elementary formalism. As before, it can be parametrized in terms of the physical fields $b^{(lm)}_{\perp}, b^{(lm)}$ with $l=1,2,\ldots$ and the harmonic fields 
$b_0^{(lm)}$ with $l=0$.

\newsec{Logarithmic Quantum Corrections: the Vector field}
In this section we compute functional determinants with the heat kernel method \refs{\GopakumarQS, \HawkingJA, \VassilevichXT}. We first review the elementary heat kernels that we need, including the basic contribution from boundary modes. We 
then compare the on-shell and off-shell computations of the heat kernel for a vector field. 

\subsec{Elementary Heat Kernels}
The basic heat kernel for a massless scalar on the sphere $S^2$ is
\eqn\fa{\eqalign{
K^s_S = {1\over 4\pi a^2}\sum_{k=0}^\infty e^{-k(k+1)s} (2k+1) &= {1\over 4\pi a^2 s} ( 1 + {1\over 3} s + {1\over 15}s^2+\ldots )~.
}}
Each component of a vector field on $S^2$ has the same spectrum as a scalar field on $S^2$ but the $k=0$ mode is absent from the partial wave expansion. Therefore the heat kernel for a vector on $S^2$ is
\eqn\faa{\eqalign{
K^v_S = {1\over 4\pi a^2}\sum_{k=1}^\infty e^{-k(k+1)s} (2k+1) &= K^s_S  -  {1\over 4\pi a^2}= {1\over 4\pi a^2 s} ( 1 - {2\over 3} s + {1\over 15}s^2+\ldots ) ~.
}}
We also need the scalar heat kernel on AdS$_2$. The representation of a heat kernel as an expansion in around flat space shows that the local terms are determined from $K^s_S$ by flipping the sign of terms that are odd in the curvature so:
\eqn\fb{
K^s_A = {1\over 4\pi a^2 s} ( 1 - {1\over 3} s + {1\over 15}s^2+\ldots )~.
}
Although this rule of thumb applies for local terms, there is no similar continuation of eigenvalues and eigenfunctions \refs{ \CamporesiGA, \CamporesiNW, \CamporesiWN}.
The heat kernels above refer to 2D fields on AdS$_2$ and $S^2$. 
We assemble these 2D heat kernels into heat kernels for 4D fields on AdS$_2 \times S^2$ 
by summing over towers of the form
\eqn\fc{
K^s_4 = K^s_A \cdot {1\over 4\pi a^2} \sum_j e^{-m^2_js} (2j+1)~,
}
where each value of angular momentum $j$ on $S^2$ has a specific value of the effective AdS$_2$ mass $m^2_j= h_j(h_j-1)$. For example, dimensional reduction of a massless 4D scalar field on $S^2$ gives a tower of 2D fields with the AdS$_2$ Casimir $h_j(h_j-1)$ identical to the $S^2$ Casimir $j(j+1)$. In this case the spectrum is $(h,j) = (k+1,k)$ with $k=0,\ldots$ so $h_j=j+1$ and the sum in \fc\ reduces to the sum in \fa. We therefore find
\eqn\fd{
K_4^s = K^s_A K^s_S 
= {1\over 16\pi^2 a^4 s^2} ( 1  +  {1\over 45}s^2+\ldots  )~.
}

The physical components arising from reduction of a 4D vector field is restricted to helicities $\pm 1$ but otherwise identical to two 4D scalar fields.  The conformal weights for a single tower of this type is therefore again $(h,j) = (k+1,k)$ but with $k=1,\ldots$ because the angular momentum $j=0$ on the $S^2$ is prohibited. The sum over $S^2$ quantum numbers reduces to \faa\ and so we find
\eqn\fe{
K_4^\prime = {1\over16\pi^2 a^4 s^2} ( 1 - {1\over 3} s + {1\over 15}s^2+\ldots )( 1 - {2\over 3} s + {1\over 15}s^2+\ldots )
= {1\over 16\pi^2 a^4 s^2} ( 1 -s +   {16\over 45}s^2+\ldots )~,
}
for a 4D scalar with partial wave $j=0$ missing. 

\subsec{Counting Boundary Modes}
The harmonic modes are zero-modes from the AdS$_2$ point of view. Their heat kernel is given by a sum over a complete set of modes that takes the schematic form 
\eqn\bqa{
K(x, x';s) = \sum_i f_i(x) f^*_i(x') ~.
}
We presented all harmonic modes in \bu. At equal points the sum over all harmonic modes for the vector field in the geometry \bt\ gives 
\eqn\bx{\eqalign{
K(x, x;s) & =\sum_{n=1}^\infty \bigg( |\nabla u_n |^2  + {\rm c.c.} \bigg)= 2\sum_{n=1}^\infty g^{z{\bar z}} \partial_z u_n \partial_{\bar z} u^*_n \cr 
& = \sum_{n=1}^\infty  {(1-r^2)^2\over a^2}  {1\over 2\pi n} n^2r^{2(n-1)} = {1\over 2\pi a^2}~. 
}}
The expression is independent of the position $r$, as expected in a homogeneous space. 

Homogeneity of AdS$_2$ allows us to write alternatively 
\eqn\bqa{\eqalign{
K(x, x; s) & =  {1\over {\rm Vol}} \int \sqrt{g} d^2 z \sum_i |f_i(x)|^2= {1\over {\rm Vol}_c} N_c~,
}}
where ${\rm Vol}_c$ is the regulated AdS$_2$ volume and $N_c$ is the regulated number of harmonic modes. Thus the equal point 
heat kernel can be interpreted as the {\it density} of harmonic modes in AdS$_2$. 

We interpret the finite density \bx\ as the contribution to the heat kernel from a single massless boundary mode rather than a field on the 1D boundary of AdS$_2$. 

\subsec{Heat Kernel for a 4D Vector Field: the Off-shell Method}
We can arrive at the heat kernel for a 4D vector field by adding contributions from all four components of the vector field and then cancel two unphysical components by introducing ghosts. This is the strategy that is most commonly used. 

In this off-shell method the two towers originating from vector components along $S^2$ are treated identically. They were denoted $b^{(lm)}, \tilde{b}^{(lm)}$ in the explicit mode expansion \ae. From the AdS$_2$ point these are towers of scalars fields with the leading partial wave $j=0$ missing so their heat kernel is given by \fe.

In the off-shell method the two towers of scalars originating from vector components along AdS$_2$ are also treated identically. They were denoted $b^{(lm)}_{\parallel}, b^{(lm)}_{\perp}$ in the explicit mode expansion. 
The direct computation of the heat kernel on AdS$_2$ requires consideration of a complete set of vector modes on AdS$_2$ and subsequent summation over the $S^2$ tower. The appropriate modes were identified in \CamporesiGA. 
For the present purpose recall that heat kernels can be represented as a local expansion. We can therefore take a short-cut and simply invert the sign of the linear term in \fe, corresponding to the interchange $A\leftrightarrow S$. This gives
\eqn\gb{
2\tilde{K}_4^\prime  = {1\over 8\pi^2 a^4 s^2} ( 1  + s +   {16\over 45}s^2+\ldots  )~.
}

The final contribution to the off-shell computation are the two ghosts \br\ which are standard scalars with heat kernel given in \fd\ except for an overall sign due to statistics. The net result for the 4D vector field then becomes
\eqn\gd{
K_4^v  = 2K_4^\prime + 2\tilde{K}_4^\prime - 2K_4^s = {1\over 8\pi^2 a^4 s^2} ( 1  +  {31\over 45}s^2+\ldots  )~.
}
%

\subsec{Heat Kernel for a 4D Vector Field: the On-shell Method}
The on-shell computation focusses on the physical components of the 4D vector field. These are two towers of scalar fields on AdS$_2$ with 
angular momentum on the $S^2$ $l=1,2,\ldots$. In our explicit mode expansions these two towers of physical modes are 
$b^{(lm)},b^{(lm)}_{\perp}$ 
with $l=1,2,\ldots$. 
They each contribute to the heat kernel with $K'_4$ given in \fe. 

In the on-shell computation the only additional contribution is a single tower of boundary modes on AdS$_2$ with partial wave expansion $l=0,1,\ldots$. There is one such mode for each of the AdS$_2$ pairs $b^{(lm)}_{\perp}, b^{(lm)}_{\parallel}$ $l=0,1,\ldots$ or, equivalently, one for each gauge function $\lambda^{(lm)}$ $l=0,1,\ldots$. For each entry in the tower the AdS$_2$ part contributes with a factor of the regulated AdS$_2$ volume with 
normalization \bx. The sum \fc\ over the $S^2$ tower of boundary modes thus contributes a simple scalar field on $S^2$ $\fa$. 

In the on-shell computation the heat kernel for the 4D vector field becomes 
\eqn\ge{\eqalign{
K_4^v  &= 2K_4^\prime+ {1\over 2\pi a^2} K^s_S \cr
&={1\over 8\pi^2 a^4 s^2} ( 1  - s +  {16\over 45}s^2 ) + {1\over 8\pi^2 a^4}  ({1\over s} +{1\over 3})\cr
&={1\over 8\pi^2 a^4 s^2} ( 1  +  {31\over 45}s^2 )~.
}}
This agrees with the off-shell result \gd.

The off-shell and the on-shell computations are related by a simple rearrangement. 
\eqn\gf{\eqalign{
K_4^v  &= 2K_4^\prime + 2\tilde{K}_4^\prime - 2K_4^s  = 2K_4^\prime 
 + 2(K^s_A + {1\over 4\pi a^2})K^s_S - 2K^s_A K^s_S\cr
&= 2K_4^\prime+ {1\over 2\pi a^2} K^s_S~. 
}}
The key is that the subtraction of the $l=0$ mode for a vector on $S^2$ included in \fe\ amounts to an addition of the boundary mode in AdS$_2$ that is implicitly included in \gb. 

Some mild virtual aspects remain in on-shell method. The heat kernel \fb\ of a bulk field in AdS$_2$ implicitly sums over the continuum of off-shell modes of plane-wave type. 
Similarly, the boundary mode has fixed wave function on AdS$_2$ but the sum over the tower of $S^2$ partial waves  probes the configuration space off-shell. The simplification of the   
on-shell computation is that we do not need to determine the explicit spectrum of the gauge violating modes, longitudinal modes, and the corresponding ghosts. It is known from the outset that these contributions must cancel so we may as well not compute them in the first place. Instead, we include just the boundary modes which appear with positive sign, as expected from physical modes.

\newsec{Supergravity in AdS$_2 \times S^2$ - Bosonic Sector}

In this section we analyze the bosonic sector of ${\cal N}=2$ supergravity in AdS$_2 \times S^2$. The matter content is a tensor field $h_{IJ}$ coupled to a vector field $a_I$.  We derive the linearized equations of motion from the AdS$_2$ point of view, then diagonalize them explicitly and find the full spectrum and all eigenvectors. Finally, we write the modes in a basis where their gauge transformations are manifest. This classifies the modes as gauge violating, pure gauge, or physical.

\subsec{4D Theory}

The 4D action for the gravity-graviphoton system is just standard Einstein-Maxwell
\eqn\ha{
{\cal L_{\rm EM}} ={1\over 2}\left[ R - {1\over 4}F_{IJ}F^{IJ}\right]~.
}
The physical content of the theory can be extracted by imposing Lorentz gauge
\eqn\hzb{\eqalign{
\nabla^I h_{IJ} -{1\over2}&\nabla_J h_{\phantom{I}I}^{I}=0~, \cr
\nabla^I a_I &= 0~,
}}
on the perturbations $\delta g_{IJ}= h_{IJ}$, $\delta A_I = a_I$. We once again implement this in the quantum theory by adding gauge fixing terms to the action and taking Feynman gauge $\xi =1$. The gauge fixed action is
\eqn\hc{
{\cal L_{\rm Feynman}} ={1\over 2}\left[ R - {1\over 4}F^2 -{1\over 2} \bigg(\nabla^I h_{IJ} -{1\over2}\nabla_J h_{\phantom{I}I}^{I}\bigg)^2 -{1\over 2}(\nabla^I a_I )^2\right]~.
}

We consider the magnetic AdS$_2\times S^2$ background. In our units the background reads 
\eqn\hea{
F_{\alpha\beta} = 2a\epsilon_{\alpha\beta}~, R_{\mu\nu} = - a^{-2} g_{\mu\nu}~, R_{\alpha\beta} = a^{-2} g_{\alpha\beta}~.
}
We take the scale $a=1$ in this section but restore it later. 

When analyzing the spectator vector field in AdS$_2\times S^2$ we diagonalized the 4D action before reducing it on $S^2$.  In the present context it is simpler to take the linearized equations of motion in 4D, reduce them on $S^2$, and only then diagonalize. We therefore first consider the gauge fixed Maxwell's equations in 4D:
\eqn\he{
\nabla^I F_{IJ} + \nabla_J \nabla^I a_I = 0~.
}
Perturbing around the background \hea\ and keeping only linear terms yields
\eqn\hf{\eqalign{
 - 2\nabla^ \alpha  h_{\mu}^{\phantom{\mu } \beta}\epsilon_{\alpha \beta} + (\nabla^2_A + \nabla^2_S +1)a_\mu =0~.
}}
\eqn\hfa{\eqalign{
-  2\nabla_\mu h^{\mu\alpha}\epsilon_{\alpha \beta} + \nabla^\alpha (h_{\mu}^\mu-h_\gamma^\gamma)\epsilon_{\alpha \beta} + (\nabla^2_A + \nabla^2_S -1)a_\alpha =0~.
}}

An analogous computation for Einstein's equations yields
%
%
%
\eqn\hh{\eqalign{
- {1\over 2} (\nabla^2-2 )h_{\alpha\beta} + {1\over 4}g_{\alpha \beta} \big[ (\nabla^2 + 2)h_\gamma^\gamma + (\nabla^2 - 2)h_\rho^\rho   \big]=  g_{\alpha\beta} \epsilon^{\gamma\delta} \nabla_\gamma a_\delta~,  }}
  \eqn\hha{\eqalign{
- {1\over 2} (\nabla^2 +2) h_{\mu\nu} + {1\over 4}g_{\mu \nu} \big[ (\nabla^2 - 2)h_\gamma^\gamma + (\nabla^2 + 2)h_\rho^\rho   \big] =   - g_{\mu\nu} \epsilon^{\alpha\beta} \nabla_\alpha a_\beta~,\cr
}}
  \eqn\hhb{\eqalign{
 {1\over 2} (\nabla^2 -2) h_{\mu\alpha} =  \epsilon_{\alpha\beta} \left( \nabla_\mu a^\beta - \nabla^\beta a_\mu\right)~. }}

The graviphoton equations of motion \hf -\hfa\ are more complicated than those for a spectator vector field because here we allow the metric to fluctuate as well. Similarly, the vector field terms in \hh-\hhb\ constitute nontrivial kinetic mixing.

\subsec{2D Effective Theory.}

We want to represent the 4D equations of motion \hf-\hhb\  as towers of 2D equations.
The physics of the 2D theory is determined by Kaluza-Klein reduction in homogeneous spaces \SalamXD. As in \ae\ we expand the 4D fields in partial waves:
\eqn\ia{\eqalign{
h_{\{\mu\nu\}}(x,y) &= H^{(lm)}_{\{\mu\nu\}}(x)Y_{(lm)}(y)~, \cr
h_{\rho}^{\phantom{\rho}\rho}(x,y) &= H_{\phantom{(lm)}\rho}^{(lm)\phantom{\rho}\rho}(x)Y_{(lm)}(y)~, \cr
h_{\mu \alpha}(x,y) &= \tilde{B}_\mu^{(lm)}(x) \nabla_\alpha Y_{(lm)}(y) + B_\mu^{(lm)}(x)\epsilon_{\alpha \beta} \nabla^\beta Y_{(lm)}(y)~, \cr
h_{\{\alpha \beta \}}(x,y) &= \phi^{(lm)}(x)\nabla_{\{ \alpha}\nabla_{\beta \}}Y_{(lm)}(y) + \tilde{\phi}^{(lm)}(x)\nabla_{\{ \alpha}\epsilon_{\beta \}\gamma}\nabla^\gamma Y_{(lm)}(y)~, \cr
h_\alpha^{\phantom{\alpha}\alpha}(x,y) &= \pi^{(lm)}(x)Y_{(lm)}(y)~, \cr
a_\mu(x,y) &= b_\mu^{(lm)}(x)Y_{(lm)}(y)~, \cr
a_\alpha(x,y) &= \tilde{b}^{(lm)}(x) \nabla_\alpha Y_{(lm)}(y) + b^{(lm)}(x)\epsilon_{\alpha \beta} \nabla^\beta Y_{(lm)}(y)~.
}}
Sum over angular momentum quantum numbers $(lm)$ is implied. Curly
brackets around indices indicate that we remove the 2D trace: $h_{\{\alpha \beta \}} = h_{\alpha \beta } - {1\over2}g_{\alpha\beta}h_\gamma ^{\phantom{\gamma}\gamma}$, and analogously for $h_{\{\mu\nu\}}$, \ChristensenIY. We also expand the generators of diffeomorphisms and gauge transformations in spherical harmonics,
\eqn\ib{\eqalign{
\xi_\mu(x,y) & = \xi^{(lm)}_\mu(x) Y_{(lm)}(y)~,\cr
\xi_\alpha(x,y) & = \zeta^{(lm)} (x)\nabla_\alpha Y_{(lm)}(y) + \xi^{(lm)}(x)\epsilon_{\alpha \beta} \nabla^\beta Y_{(lm)}(y)~, \cr
\Lambda(x,y)  &= \lambda^{(lm)} (x)Y_{(lm)}(y)~.
}}
The allowed range for the angular momentum quantum number of each mode can be read off from the expressions \ia\ and \ib. The modes with a single (double) derivative acting on the spherical harmonic functions are missing the first (the first two) modes. The table below summarizes the allowed range of $l$ for all 2D modes defined in \ia\ and \ib.
\vskip 0.1in
{
\offinterlineskip
\tabskip=0pt
\halign{ 
\vrule height3.25ex depth1.25ex width 1pt #\tabskip=2em & \hfil #\hfil &\vrule width 1pt # &  \hfil #\hfil &#\vrule  width 1pt \tabskip=0pt\cr
 \noalign{ \hrule height 1.0pt}
& { \bf 2D Field; Gauge Parameter } && \omit Range & \cr
\noalign{\hrule height 1.0pt}
& $H^{(lm)}_{\{\mu\nu\}}$, $H_{\phantom{(lm)}\rho}^{(lm)\phantom{\rho}\rho}$, $\pi^{(lm)}$, $b_\mu^{(lm)}$~; $\xi^{(lm)}_\mu$, $\lambda^{(lm)}$ && $ l= 0,1...  $ &\cr \noalign{\hrule}
& $\tilde{B}_\mu^{(lm)}$, $B_\mu^{(lm)}$, $\tilde{b}^{(lm)}$, $b^{(lm)}$~; $\zeta^{(lm)} $,  $ \xi^{(lm)}$&&  $ l= 1,2... $& \cr \noalign{\hrule}
& $\phi^{(lm)}$, $\tilde{\phi}^{(lm)}$ &&$ l= 2,3... $  &\cr \noalign{\hrule}
\noalign{\hrule height 1.0pt}
}}

\vskip 0.5in

 Inserting the partial wave expansion \ia\ into the Maxwell equations \hf-\hfa\ we find
\eqn\ic{\eqalign{
\bigg( (\nabla^2_A - l(l+1) +1) b^{(lm)}_\mu(x)  -2l(l+1)B_{\mu}^{(lm)}(x) \bigg)Y_{(lm)}(y)  = 0~,
}}
\eqn\id{\eqalign{
&\bigg(  (\nabla^2_A -l(l+1))\tilde{b}^{(lm)}(x)-2\nabla^\mu B_{\mu}^{(lm)}(x) \bigg)\nabla_\alpha Y_{(lm)}(y) + \cr
 \bigg(  (\nabla^2_A -l(l+&1))b^{(lm)}(x) +2\nabla^\mu  \tilde{B}_{\mu}^{(lm)}(x) + \pi^{(lm)}(x) - H_{\phantom{(lm)}\rho}^{(lm)\phantom{\rho}\rho}(x) \bigg)\epsilon_{\alpha \beta} \nabla^\beta Y_{(lm)}(y)=0~.
}}
The dependence on the $S^2$ coordinates can be integrated out by contracting \ic\ and \id\ with the appropriate spherical harmonic functions and using their orthonormality conditions. The result is one equation that is a vector from the AdS$_2$ point of view and two equations that are scalars.

Dimensional reduction of the Einstein equations \hh-\hhb\ proceeds similarly. For brevity we just present a summary of all 2D effective equations of motion.

\vskip 0.1in
{\bf 2D Equations of Motion - Summary}
\vskip 0.1in

The equations defined for $l=0,1...$ are
\eqn\ica{\eqalign{
(\nabla^2_A - l(l+1) +1) b^{(lm)}_\mu  -2l(l+1)B_{\mu}^{(lm)} = 0~,
}}
\eqn\iec{\eqalign{
- {1\over 2} (\nabla^2_A -l(l+1)-2) \pi^{(lm)}  -2l(l+1) b^{(lm)} =0~,
}}
\eqn\iga{\eqalign{
-{1\over 2} (\nabla^2_A-l(l+1)+2) H_{\{\mu\nu\}}^{(lm)}  =0~,
}}
\eqn\igb{\eqalign{  -{1\over 2} (\nabla^2_A-l(l+1)-2) H_{\phantom{(lm)}\rho}^{(lm)\phantom{\rho}\rho} -2\pi^{(lm)}+2l(l+1) b^{(lm)} =0~. 
}}

The equations defined for $l=1,2...$ are
\eqn\ida{\eqalign{
  (\nabla^2_A -l(l+1))\tilde{b}^{(lm)}-2\nabla^\mu B_{\mu}^{(lm)}=0~, 
}}
\eqn\idb{\eqalign{
  (\nabla^2_A -l(l+1))b^{(lm)} +2\nabla^\mu \tilde{B}_{\mu}^{(lm)} + \pi^{(lm)} - H_{\phantom{(lm)}\rho}^{(lm)\phantom{\rho}\rho} =0~,
}}
\eqn\ifaa{\eqalign{
- {1\over 2}( \nabla^2_A-l(l+1)-1)\tilde{B}_\mu^{(lm)} + \nabla_\mu b^{(lm)}=0~,
 }}
 \eqn\ifab{\eqalign{ 
 - {1\over 2}( \nabla^2_A-l(l+1)-1)B_\mu^{(lm)} -  \nabla_\mu \tilde{b}^{(lm)}+b_\mu^{(lm)} =0~.
 }}

The equations defined for $l=2,3...$ are
\eqn\iea{\eqalign{
- {1\over 2} (\nabla^2_A -l(l+1) +2)  \phi^{(lm)} =0~,
}}
\eqn\ieb{\eqalign{ 
 - {1\over 2} (\nabla^2_A -l(l+1) +2) \tilde{\phi}^{(lm)} =0 ~.
}}

The complete set of equations has $10+4=14$ components as expected for gravity coupled to a gauge field. They are organized into $6$ scalar equations, $3$ vector equations (with two components each), and one equation that is a symmetric traceless tensor (with two components).

\subsec{Spectrum}
To compute the 2D spectrum we must diagonalize the system of 2D equations of motion presented above. To disentangle the equations we dualize each of the 2D vectors $B_\mu^{(lm)}, \tilde{B}_{\mu}^{(lm)}, b_\mu^{(lm)}$ into two scalars and one harmonic mode, as in \be. A new feature is that we also need to dualize the symmetric traceless tensor $H_{\{\mu\nu\}}^{(lm)}$ to scalars \GibbonsJI. We write
\eqn\ih{
H_{\{\mu\nu\}}^{(lm)} = \nabla_{\{\mu} \nabla_{\nu\}} H^{(lm)}_{+} + \nabla_{\{\mu} \epsilon_{\nu\}\rho}\nabla^\rho H^{(lm)}_{\times}+\nabla_{\{\mu} \nabla_{\nu\}} H^{(lm)}_{0} .
}
The configuration space of scalars $H^{(lm)}_{+},H^{(lm)}_{\times}$ could generate all possible $H_{\{\mu\nu\}}^{(lm)}$. Indeed, to avoid that some $H_{\{\mu\nu\}}^{(lm)}$ are counted twice we 
require:
\eqn\ihaa{\eqalign{
\nabla^2_A(\nabla^2_A -2) H^{(lm)}_{+} & \neq 0~, \cr
\nabla^2_A(\nabla^2_A -2)  H^{(lm)}_{\times} & \neq 0~.
}}
For those configurations that could have been represented in either $H_{+}$ or $H_{\times}$ form we introduced the harmonic mode $H^{(lm)}_{0}$, written to be definite in its $H_{+}$ form. The harmonic mode satisfies
\eqn\ihab{\eqalign{
\nabla^2_A(\nabla^2_A -2) H^{(lm)}_{0} & = 0~. 
}}
To verify these claims it is useful to first compute
\eqn\ihaf{
\nabla^\mu H^{(lm)}_{\{\mu\nu\}} = {1\over 2} \nabla_\nu (\nabla^2_A -2) (H^{(lm)}_{+} +H^{(lm)}_{0})
+ \epsilon_{\nu\mu}\nabla^\mu (\nabla^2_A -2) H^{(lm)}_{\times} ~,
}
in AdS$_2$ and then use this identity to find $H^{(lm)}_{+}, H^{(lm)}_{\times}$ in terms of $H^{(lm)}_{\{\mu\nu\}}$. The resulting expressions involve the inverse of the operator $\nabla^2_A(\nabla^2_A -2)$ which is invertible 
on the appropriate subspaces due to \ihaa.  

%
%
%
%

After dualization of all fields to scalars the equations of motion \ica-\ieb\ can be recast as $14$ Klein-Gordon equations coupled by a $14\times 14$ block diagonal mass matrix. We find that $5$ components of the mass matrix are diagonal in our basis. The remaining blocks in the equations of motion are the $2\times 2$ block,
\eqn\ihag{\eqalign{
&(\nabla^2_A - l(l+1)) \pmatrix{ B_{\perp}^{(lm)}  \cr  b^{(lm)}_{\perp}} =\pmatrix{ 2 & 2 \cr 2l(l+1) & 0 } \pmatrix{ B_{\perp}^{(lm)}  \cr  b^{(lm)}_{\perp}}~,
  }}
the $3\times 3$ block, 
\eqn\ihah{\eqalign{
(\nabla^2_A - l(l+1))\pmatrix{B_{\parallel}^{(lm)}  \cr  b^{(lm)}_{\parallel} \cr \tilde{b}^{(lm)}} = \pmatrix{ 2 & 2 & -2 \cr 2l(l+1)& 0 & 0 \cr 4+2l(l+1) & 4 & -4 }\pmatrix{ B_{\parallel}^{(lm)}  \cr  b^{(lm)}_{\parallel} \cr \tilde{b}^{(lm)}}~,
  }}
and the $4\times 4$ block 
\eqn\ihai{\eqalign{
(\nabla^2_A - l(l+1))\pmatrix{ H^{(lm)} \cr \tilde{B}_{\parallel}^{(lm)} \cr \pi^{(lm)} \cr b^{(lm)}} =  \pmatrix{ 2 & 0 & -4 &  4l(l+1)\cr 0 & 2 & 0& 2\cr 0& 0& 2& -4l(l+1) \cr 1 & -2(2 + l(l+1)) & -1 & - 4} \pmatrix{ H^{(lm)} \cr \tilde{B}_{\parallel}^{(lm)} \cr \pi^{(lm)} \cr b^{(lm)}}~.
}}
The final $4\times 4$ block is the most complicated with eigenvectors 
\eqn\iia{\eqalign{
V_0 &= 2l(l+1)\tilde{B}^{(lm)}_{\parallel}+\pi^{(lm)}~, \cr
V_1 &= H_{\phantom{(lm)}\rho}^{(lm)\phantom{\rho}\rho}-2(2+l(l+1))\tilde{B}^{(lm)}_{\parallel}\ -{l-1\over l+1}\pi^{(lm)}-2(l+1)b^{(lm)}~, \cr
V_2 &= -lH_{\phantom{(lm)}\rho}^{(lm)\phantom{\rho}\rho}+2(2+l(l+1))l \tilde{B}^{(lm)}_{\parallel}+(l+2)\pi^{(lm)}-2l^2b^{(lm)}~, \cr
V_3 &= - H_{\phantom{(lm)}\rho}^{(lm)\phantom{\rho}\rho}+2(2+l(l+1))\tilde{B}^{(lm)}_{\parallel}+4b^{(lm)}~.
}}
Our result for the spectrum and the corresponding modes is:

\vskip 0.5in
\centering=10em
{
\offinterlineskip
\tabskip=0pt
\halign{ 
\vrule height3.25ex depth1.25ex width 1pt #\tabskip=2em & \hfil #\hfil &\vrule width 1pt  # & \hfil#\hfil &\vrule width 1pt# &  \hfil #\hfil &#\vrule  width 1pt \tabskip=0pt\cr
 \noalign{ \hrule height 1.0pt}
& {\bf Mode } && Mass && \omit Range & \cr
\noalign{\hrule height 1.0pt}
& $H^{(lm)}_{+}$&&   $m^2= l(l+1)+2 $ &&$ l=  0, 1...$&\cr \noalign{\hrule}
& $H^{(lm)}_{\times} $&&   $m^2= l(l+1)+2 $ &&$ l=  0, 1...$&\cr \noalign{\hrule}
& $V_0 = 2l(l+1)\tilde{B}^{(lm)}_{\parallel}+\pi^{(lm)}$ $\#$ &&$m^2=l(l+1) +2 $ &&$ l= 0, 1 ... $ &\cr \noalign{\hrule}
& $\tilde{B}^{(lm)}_{\perp} $ && $m^2=  l(l+1) +2 $ && $ l= 1, 2... $ &\cr \noalign{\hrule}
& $b_{\perp}^{(lm)} -l B^{(lm)}_{\perp}$ && $m^2= l(l-1)$ && $ l= 0, 1...  $ &\cr \noalign{\hrule}
& $V_1 $ && $m^2= l(l-1)$&& $ l= 1, 2 ... $& \cr \noalign{\hrule}
& $V_2$&& $m^2= (l+1)(l+2)$ &&$ l= 0,1... $  &\cr \noalign{\hrule}
& $b_{\perp}^{(lm)} +(l+1) B^{(lm)}_{\perp}$ && $m^2= (l+1)(l+2)$ &&$ l= 1, 2 ...  $  &\cr \noalign{\hrule}
& $\tilde{b}^{(lm)} -  b_{\parallel}^{(lm)} -2B^{(lm)}_{\parallel} $ $\#$ && $m^2= l(l+1)$&&$ l=  0,1, ...$  &\cr \noalign{\hrule}
&$b_{\parallel}^{(lm)}  + l(l+1)B^{(lm)}_{\parallel} ~~~\ddagger$ && $m^2= l(l+1)$  &&  $ l= 1,2 ... $ &\cr \noalign{\hrule}
& $ B^{(lm)}_{\parallel}+b^{(lm)}_{\parallel}-\tilde{b}^{(lm)}$ && $m^2= l(l+1)-2$ &&  $ l= 1,2 ... $ &\cr \noalign{\hrule}
& $\phi^{(lm)} $ && $m^2= l(l+1) -2$ &&$ l= 2, 3...  $ &\cr \noalign{\hrule}
& $ \tilde{\phi}^{(lm)}$ && $m^2=  l(l+1) -2$ &&$ l= 2, 3... $ &\cr \noalign{\hrule}
& $ V_3$ $\dagger$ && $m^2= l(l+1) -2$ &&$ l= 1, 2...  $  &\cr \noalign{\hrule}
\noalign{\hrule height 1.0pt}
}}
\vskip 0.5in

Comments:
\itemitem{$\bullet$} 
The eigenvectors $V_n$ with $n=0,1,2,3$ were defined in \iia. 
\itemitem{$\bullet$} 
We express our results for the eigenvalues as scalar masses defined in the usual way
\eqn\ii{\eqalign{
(-\nabla^2_A + &m^2) X = 0~.
}}

\itemitem{$\bullet$} We do not indicate the harmonic modes explicitly. In the present context they can be absorbed in $\parallel$ components and $+$ components. 

\itemitem{$\bullet$} The mode labeled with $ \dagger$ does not apply for $l=1$ and the two modes labelled with $\#$ similarly do not apply at $l=0$. We inspect these special cases later. 

\itemitem{$\bullet$} The entry labeled with $\ddagger $ is not a true eigenvector. Instead it is a generalized eigenvector associated with a repeated eigenvalue. We discuss the details of this issue in Appendix A.

 \subsec{Gauge Violating, Longitudinal, and Physical States}
 At this point we have diagonalized the gauge fixed equations of motion but we did not yet analyse gauge symmetry. To do so we first write the gauge conditions \hzb\ in components
 
 \eqn\ij{ \nabla^\mu h_{\{\mu \nu\}}  +  \nabla^\alpha h_{\alpha \nu} - {1\over 2}\nabla_\nu h^\alpha_\alpha =0~,
 } 
 \eqn\ija{ \nabla^\alpha h_{\{\alpha \beta\}}  + \nabla^\mu h_{\mu \beta} -  {1\over 2}\nabla_\beta h^\mu_\mu  =0~,
 } 
\eqn\ijb{
\nabla^\mu a_\mu + \nabla^\alpha a_\alpha = 0~,
}
and then insert the partial wave expansion \ia\ to find the 2D version of the gauge conditions
\eqn\ik{
\nabla^\mu H^{(lm)}_{\{ \mu\nu\}} - l(l+1)\tilde{B}^{(lm)}_{\nu} - {1\over 2}\nabla_\nu \pi^{(lm)} = 0~,
}
\eqn\ikc{
\nabla^\mu b_\mu^{(lm)} - l(l+1)\tilde{b} =0~.
}
\eqn\ika{
l(l+1)\bigg[ \nabla^\mu \tilde{B}^{(lm)}_\mu + {1\over 2}(2-l(l+1))\phi^{(lm)} - {1\over 2}H_{\phantom{(lm)}\rho}^{(lm)\phantom{\rho}\rho}\bigg] = 0~,
}
\eqn\ikb{
l(l+1)\bigg[ \nabla^\mu B^{(lm)}_\mu + {1\over 2}(2-l(l+1))\tilde{\phi}^{(lm)} \bigg] = 0~.
}
The factors of $l(l+1)$ in front of \ika\ and \ikb\ are due to the integration over the $S^2$ coordinates. We retained them to stress that these equations apply only for $l\geq 1$. The field components that are only defined at $l\geq 1$ similarly appear with a prefactor $l(l+1)$ so that the $l=0$ component is not needed; and the fields $\phi^{(lm)}$, $\tilde{\phi}^{(lm)}$ that are defined only for $l\geq 2$ both have a prefactor that vanishes at $l=0, 1$.

Our next step is to dualize the 2D vectors and the 2D tensor using \be\ and \ih. The gauge conditions defined for $l=0,1,...$ become
\eqn\il{\eqalign{
\nabla_\nu & \bigg[{1\over 2} (\nabla^2_A -2)H^{(lm)}_+ - l(l+1)\tilde{B}^{(lm)}_{\parallel} - {1\over 2} \pi^{(lm)}\bigg] \cr 
&+ \epsilon_{\nu\mu }\nabla^\mu \bigg[{1\over 2} (\nabla^2_A -2)H^{(lm)}_\times- l(l+1)\tilde{B}^{(lm)}_{\perp} \bigg] = 0~,
}}
\eqn\ila{
\nabla^2_A b_ \parallel ^{(lm)} - l(l+1)\tilde{b} =0~,
}
and those defined for $l=1,2,...$ become
\eqn\ilb{
\nabla^2_A \tilde{B}^{(lm)}_\parallel + {1\over 2}(2-l(l+1))\phi^{(lm)} - {1\over 2}H_{\phantom{(lm)}\rho}^{(lm)\phantom{\rho}\rho} = 0~,
}
\eqn\ilc{
 \nabla^2_A B^{(lm)}_\parallel + {1\over 2}(2-l(l+1))\tilde{\phi}^{(lm)} = 0~,
}
We can project \il\ and obtain two linearly independent scalar equations by applying $\nabla^\nu$ or $\epsilon^{\nu\mu }\nabla_\mu$ and then inverting the resulting overall Laplacian $\nabla^2_A$. Our results below will indeed justify the inversion except for the special case $l=0$ which we reconsider later. With this exception we can therefore simply require both square brackets in \il\ to vanish. 

Our final step is to eliminate the kinetic operators $\nabla^2_A$ from \il\ - \ilc\ by using the equations of motion. This gives the on-shell gauge conditions:

\eqn\ha{
{1\over2}l(l+1)H^{(lm)}_{+} - l(l+1)\tilde{B}^{(lm)}_\parallel - {1\over 2} \pi^{(lm)}=0~,
 }
\eqn\haa{
{1\over2}H^{(lm)}_{\times} - \tilde{B}^{(lm)}_\perp =0~,
 } 
\eqn\hd{
 \tilde{b}^{(lm)}- b_\parallel^{(lm)} - 2B_\parallel^{(lm)} =0~,
 } 
\eqn\hb{
(2+l(l+1))\tilde{B}_{\parallel}^{(lm)}+2b^{(lm)} + {1\over2}(2-l(l+1))\phi^{(lm)} - {1\over2}H_{\phantom{(lm)}\rho}^{(lm)\phantom{\rho}\rho}=0~,
 } 
\eqn\hc{
 (2+l(l+1))B_{\parallel}^{(lm)} -2\tilde{b}^{(lm)}+2b_\parallel^{(lm)}+ {1\over2}(2-l(l+1))\tilde{\phi}^{(lm)}=0~.
 }
As mentioned above, these equations apply only for $l\geq 1$ and we return to $l=0$ later.
 
The modes presented in Section 4.3 were identified only by their eigenvalues so we can freely choose a new basis by taking linear combinations of modes with the same mass. The gauge conditions \ha-\hc\ specify particular linear combinations that are set to zero by the gauge conditions. We collect these gauge violating modes in a table. 
 
 \vskip 0.5in
 
 \centering=10em
{
\offinterlineskip
\tabskip=0pt
\halign{ 
\vrule height3.25ex depth1.25ex width 1pt #\tabskip=0.5em & \hfil #\hfil &\vrule width 1pt  #  &  \hfil #\hfil &#\vrule  width 1pt \tabskip=0pt\cr
 \noalign{ \hrule height 1.3pt}
& {\bf Gauge Violating Modes} && Mass & \cr
\noalign{\hrule height 1.0pt}
& ${1\over2}l(l+1)H^{(lm)}_{+} - l(l+1)\tilde{B}^{(lm)}_\parallel - {1\over 2} \pi^{(lm)}$&&   $m^2= l(l+1) +2 $ &\cr \noalign{\hrule}
& ${1\over2}H^{(lm)}_{\times} - \tilde{B}^{(lm)}_\perp$&&   $m^2= l(l+1) +2 $ &\cr \noalign{\hrule}
& $\tilde{b}^{(lm)} -  b_{\parallel}^{(lm)} -2B^{(lm)}_{\parallel} $ && $m^2= l(l+1)$&\cr \noalign{\hrule}
& $(2+l(l+1))\tilde{B}^{(lm)}_{\parallel}+2b^{(lm)} + {1\over2}(2-l(l+1))\phi^{(lm)} - {1\over2}H_{\phantom{(lm)}\rho}^{(lm)\phantom{\rho}\rho}$ && $m^2= l(l+1) -2$ &\cr \noalign{\hrule }
& $ (2+l(l+1))B_{\parallel}^{(lm)} -2\tilde{b}^{(lm)}+2b_\parallel^{(lm)}+ {1\over2}(2-l(l+1))\tilde{\phi}^{(lm)}$ && $m^2= l(l+1)-2$ &\cr \noalign{\hrule}
\noalign{\hrule height 1.3pt}
}}

\vskip 0.5in
 
Our next step is to take equivalences under gauge and diffeomorphism transformations into account. The variations of the 4D fields are:
\eqn\he{\eqalign{
\delta a_I &= \nabla_I \Lambda' + \xi^JF_{JI} + \nabla_I (\xi^J A_J)~, \cr
\delta h_{IJ} &= \nabla_I \xi_J + \nabla_J \xi_I~.
}}
The gauge field varies under diffeomorphisms but the metric fluctuations do not vary under gauge transformations. It is therefore advantageous to remove field components in a specific order: first exploit diffeomorphisms and then gauge transformations. In particular, we have not yet specified a gauge for the background gauge fields $A_J$ although the field strength is of course specified in \hea. We take this into account by redefining diffeomorphisms to include a compensating gauge transformation that removes the $A_J$ dependence. We implement this by henceforth taking $\Lambda^\prime = \Lambda - \xi^J A_J$ in \he. 

In our on-shell approach we already fixed the gauge in \hzb\ so at this point we can focus on residual symmetries. The gauge variations \he\ that preserve the gauge conditions \hzb\  satisfy
\eqn\heaa{\eqalign{
&\nabla^2_4 \Lambda + 2 \epsilon_{\alpha \beta} \nabla^\beta  \xi^\alpha =0~, \cr
&(g_{IJ}\nabla^2_4  + R_{IJ})\xi^I = 0~.
}}
Upon expansion in partial waves \ib\ we find the 2D equations of motion for the residual symmetries. The 2D diffeomorphisms $\xi^{(lm)}_\parallel, \xi^{(lm)}_\perp$ have mass $m^2=l(l+1)+2$ and range $l=0,1,\ldots$, the $S^2$ diffeomorphisms $\zeta^{(lm)}, \xi^{(lm)}$ have mass $m^2=l(l+1)-2$ and range $l=1,2,\ldots$ while the gauge symmetry is an eigenvector satisfying
\eqn\heab{\eqalign{
(\nabla^2_A - l(l+1) )\lambda^{(lm)} - 2l(l+1) \xi^{(lm)} = 0~~~~~,~l=0,1,\ldots~. 
}}
We need only consider diffeomorphisms and gauge transformations that satisfy their appropriate on-shell condition. 

Inserting the partial wave expansions \ia\ and \ib\ into the 4D symmetry variations \he\ we find variations of all 2D fields. After complete dualization to scalars the result is

\vskip 0.5in
\centering=10em
{
\offinterlineskip
\tabskip=0pt
\halign{ 
\vrule height3.25ex depth1.25ex width 1pt #\tabskip=2em & \hfil #\hfil &\vrule width 1pt  # & \hfil#\hfil &\vrule width 1pt# &  \hfil #\hfil &#\vrule  width 1pt \tabskip=0pt\cr
 \noalign{ \hrule height 1.0pt}
& { \bf Mode} && Symmetry Variation && \omit Range & \cr
\noalign{\hrule height 1.0pt}
& $H_{\phantom{(lm)}\rho}^{(lm)\phantom{\rho}\rho}$&&   $2\nabla^2_A \xi_\parallel^{(lm)} $ &&$ l=  0, 1...$&\cr \noalign{\hrule}
& $H^{(lm)}_{+}$&&   $2\xi_\parallel^{(lm)} $ &&$ l=  0, 1...$&\cr \noalign{\hrule}
& $H^{(lm)}_{\times}$ &&$2\xi_\perp^{(lm)} $  &&$ l= 0, 1 ... $ &\cr \noalign{\hrule}
& $\tilde{B}_\parallel^{(lm)} $ && $ \xi^{(lm)}_\parallel + \zeta^{(lm)} $ &&$ l= 1, 2... $ &\cr \noalign{\hrule}
& $ \tilde{B}_\perp^{(lm)} $ && $\xi^{(lm)}_\perp$  && $ l= 1, 2...  $ &\cr \noalign{\hrule}
& $B_\parallel^{(lm)}$ && $\xi^{(lm)} $ &&$ l= 1, 2 ... $& \cr \noalign{\hrule}
& $B_\perp^{(lm)}$&& $0$  &&$ l= 1, 2... $  &\cr \noalign{\hrule}
& $\phi^{(lm)} $ && $2\zeta^{(lm)}$  &&$ l=  2, 3...$  &\cr \noalign{\hrule}
& $ \tilde{\phi}^{(lm)} $ && $ 2\xi^{(lm)}$  &&$ l=  2, 3... $  &\cr \noalign{\hrule}
&$ \pi^{(lm)}  $ && $-2l(l+1)\zeta^{(lm)}$  && $ l=  0, 1 ...$ &\cr \noalign{\hrule}
& $b_\parallel^{(lm)} $ && $\lambda^{(lm)}$ &&$ l=  0, 1 ...  $ &\cr \noalign{\hrule}
& $ b_\perp^{(lm)}$ && $0$  &&$ l=  0, 1 ... $ &\cr \noalign{\hrule}
& $b^{(lm)}$ && $-2\zeta^{(lm)}$  &&$ l= 1, 2...  $  &\cr \noalign{\hrule}
& $\tilde{b}^{(lm)}$ && $2\xi^{(lm)}+\lambda^{(lm)}$  &&$ l= 1, 2...  $  &\cr \noalign{\hrule}
\noalign{\hrule height 1.0pt}
}}
\vskip 0.5in

The five towers of gauge violating modes identified in \ha-\hc\ are all invariant under symmetry variations as they should be. To obtain the longitudinal states we consider our original list of 14 towers and constrain it with the gauge conditions. For example, condition \haa\ allows us to work only with $\tilde{B}^{(lm)}_\perp$ and not worry about $H^{(lm)}_{\times}$ since these fields are proportional after imposing gauge conditions. After constraining the modes in this way we find combinations that are pure gauge,

 \vskip .5in

\centering=10em
{
\offinterlineskip
\tabskip=0pt
\halign{ 
\vrule height3.25ex depth1.25ex width 1pt #\tabskip=0.5em & \hfil#\hfil &\vrule width 1pt  # & \hfil#\hfil &\vrule width 1pt# &  \hfil #\hfil &#\vrule  width 1pt \tabskip=0pt\cr
 \noalign{ \hrule height 1.3pt}
& {\bf Longitudinal Mode} && Symmetry variation  && Mass  & \cr
\noalign{\hrule height 1.0pt}
& $2l(l+1)\tilde{B}^{(lm)}_\parallel + \pi^{(lm)}$&&  $2l(l+1)\xi_\parallel^{(lm)}$   &&$m^2= l(l+1) +2 $ &\cr \noalign{\hrule}
& $\tilde{B}^{(lm)}_\perp$&& $\xi_\perp^{(lm)}$  && $m^2= l(l+1) +2 $  &\cr \noalign{\hrule}
&$ b_{\parallel}^{(lm)}  +  l(l+1)B^{(lm)}_{\parallel} $ && $\lambda^{(lm)} + l(l+1)\xi^{(lm)}$   && $m^2= l(l+1)$ &\cr \noalign{\hrule}
& $\tilde{\phi}^{(lm)}$ && $2\xi^{(lm)}$ && $m^2= l(l+1)-2$ &\cr \noalign{\hrule}
& $\phi^{(lm)} $ && $ 2\zeta^{(lm)}$  &&$m^2= l(l+1) -2$ &\cr \noalign{\hrule }
\noalign{\hrule height 1.3pt}
}}

\vskip .5in
  
The mode $b_{\parallel}^{(lm)}  +  l(l+1)B^{(lm)}_{\parallel} $ was a generalized eigenvector prior to gauge fixing. However, the state with mass $l(l+1)$ with which it was degenerate was removed by the gauge  condition \hd\ and thus $b_{\parallel}^{(lm)}  +  l(l+1)B^{(lm)}_{\parallel} $ is now a true eigenvector. 
Its symmetry variation $\lambda^{(lm)} + l(l+1)\xi^{(lm)}$ is not diagonal but in view of \heab\ it is precisely the combination that is on-shell with mass so $m^2=l(l+1)$. 
  
There is significant ambiguity in the form of the longitudinal modes we identify. We can freely add modes proportional to the gauge violating modes since those are themselves invariant under on-shell gauge transformations. Similarly (and perhaps more relevant) we can add modes proportional to the gauge invariant physical states identified below. 
  
 \vskip .5in

 After removal of five towers of gauge violating modes and five towers of longitudinal modes there remain four towers of fields that satisfy the gauge condition and cannot be represented as pure gauge states. These are the physical states. Simplifying the modes from our $14$ original towers using the gauge conditions and then forming gauge invariant combinations we find:  
 \vskip 0.5in
 
 \centering=10em
{
\offinterlineskip
\tabskip=0pt
\halign{ 
\vrule height3.25ex depth1.25ex width 1pt #\tabskip=2em & \hfil #\hfil &\vrule width 1pt  # & \hfil#\hfil &\vrule width 1pt# &  \hfil #\hfil &#\vrule  width 1pt \tabskip=0pt\cr
 \noalign{ \hrule height 1.0pt}
& {\bf Physical Modes.} &&  Mass   && \omit Range & \cr
\noalign{\hrule height 1.0pt}
& $b_{\perp}^{(lm)} - lB_{\perp}^{(lm)}$&&  $m^2= l(l-1)$ && $l=2,\ldots$  &\cr \noalign{\hrule}
& $\pi^{(lm)}+2(l+1)b^{(lm)} + (l+1)(l+2)\phi^{(lm)}$ &&   $m^2= l(l-1) $ && $l=2,\ldots$ &\cr \noalign{\hrule}
& $ \pi^{(lm)}-2lb^{(lm)} + l(l-1)\phi^{(lm)} $ && $m^2= (l+1)(l+2)$&&   $l=1,\ldots$&\cr \noalign{\hrule }
& $ b_{\perp}^{(lm)} + (l+1)B_{\perp}^{(lm)}$ && $m^2= (l+1)(l+2)$  && $l=1,\ldots$ &\cr \noalign{\hrule}
\noalign{\hrule height 1.3pt}
}}

\vskip 0.5in

The second line is just  $-{l+1\over l-1} V_1$, while the third line is ${1\over l+2} V_2$.

\subsec{l=1 modes}
Some of our results warrant special comment for small values of $l$. In this subsection we reconsider $l=1$ and 
in the next we consider $l=0$. 

There are several issues for $l=1$:
\itemitem{$\bullet$} 
The part of the 4D graviton that is a symmetric traceless tensor on $S^2$ vanishes identically for $l=1$. Consequently the modes $\phi^{(lm)} $ and $\tilde{\phi}^{(lm)} $ are only defined for $l\geq 2$. This leaves $12$  2D scalar modes at $l=1$. 
\itemitem{$\bullet$} 
For $l=1$ the eigenvalue $m^2=l(l-1)$ of $V_1$ coincides with $m^2=l(l+1)-2$ of $V_3$. In fact, $V_3=-V_1$ for $l=1$ so in this case our set of modes is incomplete in its generic form. We address this by introducing a  generalized eigenvector $V'_3 = 4b^{(1m)} + \pi^{(1m)}$ which is acted on as $\nabla^2_A V_3'=4V_1$. 
\itemitem{$\bullet$} 
We have dualized all 2D fields fully to 2D scalars. This can lead to overcounting in case of harmonic fields, which we define as those fields where $m^2=0$ after dualization of 2D vectors and those where $m^2=0$ or $m^2=2$ after dualization of 2D symmetric traceless tensors. There are no modes of this type for $l\geq 2$ but they are present for $l=0,1$. We must therefore revisit dualization.  

We present for convenience the spectrum and the corresponding modes for $l=1$: 
\vskip 0.5in
\centering=10em
{
\offinterlineskip
\tabskip=0pt
\halign{ 
\vrule height3.25ex depth1.25ex width 1pt #\tabskip=2em & \hfil #\hfil &\vrule width 1pt# &  \hfil #\hfil &#\vrule  width 1pt \tabskip=0pt\cr
 \noalign{ \hrule height 1.0pt}
& { \bf Modes}   && Mass & \cr
\noalign{\hrule height 1.0pt}
& $V_2=-H_{\phantom{(1m)}\rho}^{(1m)\phantom{\rho}\rho}+8\tilde{B}^{(1m)}_{\parallel}+3\pi^{(1m)}-2b^{(1m)}$&& $m^2= 6$ &\cr \noalign{\hrule}
& $b^{(1m)}_{\perp} + 2B^{(1m)}_{\perp} $ && $m^2= 6$ &\cr \noalign{\hrule}
& $H^{(1m)}_{+}$&&   $m^2= 4 $ &\cr \noalign{\hrule}
& $H^{(1m)}_{\times} $&&   $m^2= 4 $ &\cr \noalign{\hrule}
& $ V_0 = 4\tilde{B}^{(1m)}_{\parallel}+\pi^{(1m)}$ &&$m^2=4 $ &\cr \noalign{\hrule}
& $\tilde{B}^{(1m)}_{\perp} $ && $m^2=  4 $ &\cr \noalign{\hrule}
& $b^{(1m)}_{\parallel} + 2B^{(1m)}_{\parallel}  -\tilde{b}^{(1m)} $ && $m^2=2$&\cr \noalign{\hrule}
&$b^{(1m)}_{\parallel} + 2B^{(1m)}_{\parallel} ~~~\ddagger $&& $m^2= 2$  &\cr \noalign{\hrule}
& $b^{(1m)}_{\parallel} + B^{(1m)}_{\parallel}-\tilde{b}^{(1m)}$&& 
$m^2= 0$ &\cr \noalign{\hrule}
& $b^{(1m)}_{\perp} - B^{(1m)}_{\perp} $&& 
$m^2= 0$  &\cr \noalign{\hrule}
& $V_1=H_{\phantom{(1m)}\rho}^{(1m)\phantom{\rho}\rho}-8\tilde{B}^{(1m)}_{\parallel}\ -4b^{(1m)} $ && $m^2=0$& \cr \noalign{\hrule}
& $V'_3 = 4b^{(1m)} + \pi^{(1m)}~~~\ddagger$  && $m^2= 0$ &\cr \noalign{\hrule}
\noalign{\hrule height 1.0pt}
}}
\vskip 0.5in

The modes labeled with $\ddagger $ are generalized eigenvectors. The $m^2=2$ mode is just the $l=1$ version of the generalized state $b_{\parallel}^{(1m)}  + l(l+1)B^{(1m)}_{\parallel} $ already present for $l\geq 2$. $V'_3 = 4b^{(1m)} + \pi^{(1m)}$ is the mode particular to $l=1$ that was discussed above. 

As we have stressed we must take care not to overcount the modes with $m^2=0$ that arise from dualization of a 2D vector to a 2D scalar. In order to illuminate the issue that may arise we consider the coupled system of 
$B^{(1m)}_{\mu}$, $b^{(1m)}_{\mu}$, and $\tilde{b}^{(1m)}$ prior to dualization. The equations of motion \ica, \ida, and \ifab\ can be presented as
\eqn\hec{
(\nabla^2_A + 1) (b^{(1m)}_\mu-B^{(1m)}_\mu)= 2\nabla_\mu \tilde{b}^{(1m)} ~,
}
\eqn\hed{
(\nabla^2_A- 5) (b^{(1m)}_\mu + 2B^{(1m)}_\mu) = -4 \nabla_\mu \tilde{b}^{(1m)} ~,
}
\eqn\hee{
(\nabla^2_A -2) \tilde{b}^{(1m)} = 2\nabla^\mu B^{(1m)}_\mu~.
}
Upon dualization to 2D scalars the right hand side of \hec\ is manifestly longitudinal so for the perpendicular component $(b^{(1m)}_\perp-B^{(1m)}_\perp)$ only the left hand side remains. Taking the curvature terms into account we find that this mode is massless, as indicated in the table. However, recall that in \bd\ we explicitly defined dualization of a 2D vector such that dual components do not satisfy the harmonic condition. This mode is therefore disallowed except if the longitudinal mode $(b^{(1m)}_\parallel-B^{(1m)}_\parallel)$ is massless as well. In that event the two modes are interpreted together as a single harmonic mode. This harmonic mode forces vanishing $\tilde{b}^{(1m)}$ and this in turn decouples the vector mode $(b^{(1m)}_\mu + 2B^{(1m)}_\mu)$. We interpret the massless $(b^{(1m)}_\perp-B^{(1m)}_\perp)$ as a harmonic mode in this strong sense. 

We next consider the gauge conditions at $l=1$

 \eqn\hf{
H^{(1m)}_{+} - 2\tilde{B}^{(1m)}_\parallel - {1\over 2} \pi^{(1m)}=0~,
 }
  \eqn\hg{
{1\over 2}H^{(1m)}_{\times} - \tilde{B}^{(1m)}_\perp =0~,
 }
    \eqn\hj{
b^{(1m)}_\parallel + 2B^{(1m)}_\parallel - \tilde{b}^{(1m)}  =0~.
 }
 \eqn\hh{
4\tilde{B}_{\parallel}^{(1m)}+2b^{(1m)} - {1\over2}H_{\phantom{(1m)}\rho}^{(1m)\phantom{\rho}\rho}=0~,
 }
  \eqn\hi{
\nabla^\mu B^{(1m)}_{\mu}=0~.
 }
 
With the exception of \hi, these are the continuations to $l=1$ of the higher $l$ conditions \ha-\hc. The derivation of \hi\ is different from the one of \hc\ only in that the equations of motion were not used to simplify it so we simply revert to \ikb. 

If we proceed to dualize the gauge condition \hi\ we find that $B^{(1m)}_{\parallel}$ is harmonic which we have disallowed. Thus $B^{(1m)}_{\parallel} =0$ and so the condition \hj\ becomes a condition on the massless mode $b^{(1m)}_{\parallel} + B^{(1m)}_{\parallel}-\tilde{b}^{(1m)}$ in addition to the massive mode $b^{(1m)}_{\parallel} + 2B^{(1m)}_{\parallel}-\tilde{b}^{(1m)}$. 

On the other hand we may dualize $B^{(1m)}_{\mu}$ to the true harmonic mode that is shared between $B^{(1m)}_{\parallel}$ and $B^{(1m)}_{\perp}$. This mode satisfies the gauge condition since in this sector we have the constraint
$(b^{(1m)}_\mu + 2B^{(1m)}_\mu)=0$ and so $B^{(1m)}_\mu$ has vanishing divergence as well as vanishing curl. 

 \vskip 0.5in
 
 \centering=10em
{
\offinterlineskip
\tabskip=0pt
\halign{ 
\vrule height3.25ex depth1.25ex width 1pt #\tabskip=2em & \hfil #\hfil &\vrule width 1pt  #  &  \hfil #\hfil &#\vrule  width 1pt \tabskip=0pt\cr
 \noalign{ \hrule height 1.3pt}
& {\bf Gauge Violating Modes} && Mass & \cr
\noalign{\hrule height 1.0pt}
& $H^{(1m)}_{+} - 2\tilde{B}^{(1m)}_\parallel - {1\over 2} \pi^{(1m)}$&&   $m^2= 4 $ &\cr \noalign{\hrule}
& $H^{(1m)}_{\times} - 2\tilde{B}^{(1m)}_\perp $&&   $m^2= 4 $ &\cr \noalign{\hrule}
& $b^{(1m)}_{\parallel} + 2B^{(1m)}_{\parallel} - \tilde{b}^{(1m)} $ && $m^2= 2$&\cr \noalign{\hrule}
& $4\tilde{B}_{\parallel}^{(1m)}+2b^{(1m)} - {1\over2}H_{\phantom{(1m)}\rho}^{(1m)\phantom{\rho}\rho}$ && $m^2= 0$ &\cr \noalign{\hrule }
& $ b^{(1m)}_{\parallel} + B^{(1m)}_{\parallel}-\tilde{b}^{(1m)} $ && $m^2= 0$&\cr \noalign{\hrule}
\noalign{\hrule height 1.3pt}
}}

\vskip 0.5in

The $5$ towers of modes that we project out due to the gauge conditions are themselves gauge invariant. Among the remaining $7$ towers there are $5$ that we can present as pure gauge. The longitudinal modes are
 \vskip 0.5in

\centering=10em
{
\offinterlineskip
\tabskip=0pt
\halign{ 
\vrule height3.25ex depth1.25ex width 1pt #\tabskip=2em & \hfil#\hfil &\vrule width 1pt  # & \hfil#\hfil &\vrule width 1pt# &  \hfil #\hfil &#\vrule  width 1pt \tabskip=0pt\cr
 \noalign{ \hrule height 1.3pt}
& {\bf Longitudinal Modes} && Mass && Symmetry variation & \cr
\noalign{\hrule height 1.0pt}
& $4\tilde{B}^{(1m)}_\parallel + \pi^{(1m)}$&&   $m^2= 4 $ &&$4\xi_\parallel^{(1m)}$ &\cr \noalign{\hrule}
& $4\tilde{B}^{(1m)}_\perp$&&   $m^2=4 $ && $4\xi_\perp^{(1m)}$&\cr \noalign{\hrule}
&$  b^{(1m)}_{\parallel} + 2B^{(1m)}_{\parallel} $ &&  $m^2= 2$  && $\lambda^{(1m)}+ 2 \xi^{(1m)}$&\cr \noalign{\hrule}
& $4b^{(1m)} + \pi^{(1m)}$ && $m^2= 0$ &&$ -6\zeta^{(1m)}$&\cr \noalign{\hrule }
&$  b^{(1m)}_{\perp} - B^{(1m)}_{\perp} $ &&  $m^2= 0$  && $2 \xi^{(1m)}$&\cr \noalign{\hrule}
\noalign{\hrule height 1.3pt}
}}

\vskip .5in
The modes in the third and fourth line were generalized eigenvectors before gauge conditions were imposed but they are now true eigenvectors. 

The fifth line refers to the harmonic mode that can be presented either perpendicular or longitudinal form. The longitudinal form can obviously be presented as a pure diffeomorphism. However, the parameter $\xi$ is itself harmonic for $l=1$ so this symmetry can also be recast in perpendicular form. These presentations are entirely equivalent. 

The fourth and fifth line in the table both correspond to modes generated by $S^2$ diffeomorphisms (with a compensating gauge transformation to keep $\lambda^{(1m)}+ 2 \xi^{(1m)}$ fixed). Neither of these $l=1$ modes are smooth continuations of the towers that apply for larger values of $l$. The last one is the mode that is physical if it is harmonic since then it is formally pure gauge but with non-normalizable gauge function. 

The two remaining towers of modes satisfy the gauge conditions and they are not pure gauge. The gauge invariant form of these physical towers are the continuations from higher $l$: 

\vskip 0.5in

 \centering=10em
{
\offinterlineskip
\tabskip=0pt
\halign{ 
\vrule height3.25ex depth1.25ex width 1pt #\tabskip=2em & \hfil #\hfil &\vrule width 1pt  #  &  \hfil #\hfil &#\vrule  width 1pt \tabskip=0pt\cr
 \noalign{ \hrule height 1.3pt}
& {\bf Physical Modes} && Mass & \cr
\noalign{\hrule height 1.0pt}
& $\pi^{(1m)}-2b^{(1m)}$ && $m^2=6$ &\cr \noalign{\hrule }
& $ b^{(1m)}_{\perp} + 2B^{(1m)}_{\perp}$ && $m^2= 6$ &\cr \noalign{\hrule}
\noalign{\hrule height 1.3pt}
}}

\vskip 0.5in

\subsec{l=0 modes}
The $l=0$ sector is the truncation of gravity and a vector field to the spherically symmetric sector. It is instructive to analyze this sector in detail. 

Prior to any dualization the 2D field content is the 2D graviton $H_{\{\mu\nu\}}^{(00)}$, the AdS$_2$ volume mode $H_{\phantom{(00)}\rho}^{(00)\phantom{\rho}\rho}$, the $S^2$ volume mode $\pi^{(00)}$, and the 2D gauge field $b^{(00)}_{\mu}$. There is a total of $6$ component fields. The three continuous symmetries generated by gauge symmetry $\lambda^{(00)}$ and the AdS$_2$ diffeomorpisms $\xi^{(00)}_{\mu}$ are each expected to gauge one component field away and require another to vanish due to a constraint. Thus we expect no physical degrees of freedom in the $l=0$ sector. 

We first consider the equations of motion
\eqn\hia{\eqalign{
(\nabla^2_A  +1) b^{(00)}_\mu = 0~,
}}
\eqn\hl{\eqalign{
(\nabla^2_A+2)   H_{\{\mu\nu\}}^{(00)}=0~,
}}
\eqn\hk{\eqalign{
(\nabla^2_A -2) \pi^{(00)}  =0~,
}}
\eqn\hm{\eqalign{  (\nabla^2_A-2) H_{\phantom{(00)}\rho}^{(00)\phantom{\rho}\rho} +4\pi^{(00)}=0~. 
}}
There is no mixing between the gauge field $b^{(00)}_{\mu}$ and the gravity modes so we can treat them separately. 

The gauge field sector is simply 2D QED. Dualizing the scalars as in \ba\ the gauge fixed equation of motion \hia\ amounts to two harmonic equations for the dualized scalars $b^{(00)}_\parallel$ and $b^{(00)}_\perp$.
\eqn\hma{\eqalign{
\nabla^2_A b^{(00)}_\parallel =
\nabla^2_A b^{(00)}_\perp &=0~.
}}
Once again, recall that we define the scalars dual to vector fields requiring that they do not satisfy the harmonic condition \bd. Both these modes therefore vanish on shell. However, since the equations of motion coincides with the harmonic equation, the harmonic mode $b_{\mu 0}^{(00)}=\nabla_\mu b_0^{(00)}$ does in fact satisfy the equations of motion. This is special to the $l=0$ sector. 

We proceed similarly for the gravity modes described by the symmetric traceless tensor $H_{\{\mu\nu\}}^{(00)}$. We must again take extra care when dualizing. According to \ih\ we can dualize to two scalars $H_{+}^{(00)}, H_{\times}^{(00)}$ which cannot satisfy the generalized harmonic condition
\eqn\hmaa{
\nabla^2_A ( \nabla^2_A -2)X =0~, 
}
and one harmonic scalar $H_0^{(00)}$ that must satisfy this equation. 

Inserting the expansion \ih\ of $H_{\{\mu\nu\}}^{(00)}$ into \hl\ we find that the equations of motion for the two dual scalars $H_{+}^{(00)}$ and $H_{\times}^{(00)}$ are precisely the generalized harmonic condition. 
These modes must therefore must vanish on shell.

However, again we find that since the equations of motion coincide with the harmonic equation, the harmonic mode $H_{\{\mu\nu\}}^{(00)}=
\nabla_{\{\mu}\nabla_{\nu\}} H_0^{(00)}$ with $H_0^{(00)}$ satisfying \hmaa\ does in fact satisfy the equations of motion. 

The remaining two modes are $H_{\phantom{(00)}\rho}^{(00)\phantom{\rho}\rho}$ and $\pi^{(00)}$. These are already scalars so we do not have to worry about any dualization. The equations of motion \hk-\hm\ indicate that these scalars have $m^2=2$. Indeed, they are equivalent to a single ``weight-two" scalar with $m^2=2$ and satisfying 
\eqn\hkq{\eqalign{
(\nabla^2_A-2)^2 H_{\phantom{(00)}\rho}^{(00)\phantom{\rho}\rho} =0~.
}}
Either way, both these scalars remain after the gauge fixed equations of motion are imposed. 

Summarizing so far, the fields that are on-shell at $l=0$ are the harmonic scalar $b_0^{(00)}$ dual to the 2D gauge field, the generalized harmonic scalar $H_0^{(00)}$ dual to the traceless symmetric tensor, and the two scalars $H_{\phantom{(00)}\rho}^{(00)\phantom{\rho}\rho}$ and $\pi^{(00)}$ with $m^2=2$. 

The 4D gauge condition for diffeomorphisms \ik\ simplifies at $l=0$ to
\eqn\hna{
\nabla^\mu H_{\{\mu\nu\}}^{(00)} = {1\over 2} \nabla_\nu \pi^{(00)}  ~.
}
We insert \ihaf\ into \hna, giving the condition
\eqn\hnb{
\nabla^\mu (\nabla^2_A-2)H_0^{(00)} = {1\over 2} \nabla_\nu \pi^{(00)}  ~,
}
We can contract with $\nabla^\nu$ and find $\nabla^2_A \pi^{(00)} =0$ in view of the generalized harmonic condition on $H_0^{(00)}$. This conflicts with the equation of motion \hk\ so we conclude that $\pi^{(00)} =0$ after the equations of motion and the gauge condition have been imposed. Further, the gauge condition \hnb\ then projects on to the $m^2=2$ component of $H_0^{(00)}$. 

The dualization of the on-shell physical fields $H_{\{\mu\nu\}}^{(00)}$ and $b^{(00)}_\mu$ manifestly presents them as pure gauge. The AdS$_2$-volume $H_{\phantom{(00)}\rho}^{(00)\phantom{\rho}\rho}$ mode is also pure gauge with gauge function chosen such that
\eqn\ho{
H_{\phantom{(00)}\rho}^{(00)\phantom{\rho}\rho} = 2\nabla_\rho \xi^{\rho(00)}~.
}
Since $H_{\phantom{(00)}\rho}^{(00)\phantom{\rho}\rho}$ has $m^2=2$ the harmonic component of $\xi^{\rho(00)}$ can play no role here. We dualize $\xi_\rho^{(00)} = \nabla_\rho \xi_\parallel^{(00)}$ where $\xi_\parallel^{(00)}$ also has $m^2=2$ as already found in \heaa. We therefore have 
\eqn\ic{ 
H_{\phantom{(00)}\rho}^{(00)\phantom{\rho}\rho} = 2\nabla_A^2 \xi_\parallel^{(00)} =4\xi_\parallel~. 
}
on-shell. In particular, it is manifest that all normalizable $H_{\phantom{(00)}\rho}^{(00)\phantom{\rho}\rho}$ are generated by normalizable gauge functions. 

In summary, the only physical modes at $l=0$ are the harmonic modes $b_0^{(00)}$, $H_0^{(00)}$. These modes are pure gauge so we find that in this sector gauge symmetries remove all fields (at least formally). This is the expected result.  

\subsec{Boundary Modes}
As we have stressed, special care must be taken when the dualization of vector or tensor fields gives rise to harmonic modes. 

An important example of this situation is a 2D vector field that satisfies \bia\
\eqn\mb{
(\nabla^2_A+1)C_\mu=0~,
} 
since then the dual scalar field $X$ satisfies the harmonic equation $\nabla^2_A X=0$. In this case the gradient and curl versions of dualization are equivalent so only one of these configurations should be counted.   

There are three 2D vector fields in our setting. Their equations of motion simplify when we focus on harmonic fields since those are divergence free and so their couplings to gradients of scalars can be consistently ignored. With these simplifications
\ifaa\ becomes
\eqn\icb{
(\nabla^2_A - l(l+1)+1) \tilde{B}_{\mu}^{(lm)}  = 2\tilde{B}_\mu~,
}
and \ica, \ifab\ combine to 
\eqn\icf{\eqalign{
&(\nabla^2_A - l(l+1)+1) \pmatrix{ B_{\mu}^{(lm)}  \cr  b^{(lm)}_{\mu}} =\pmatrix{ 2 & 2 \cr 2l(l+1) & 0 } \pmatrix{ B_{\mu}^{(lm)}  \cr  b^{(lm)}_{\mu}}~.
  }}
We must in addition consider the 2D tensor $H^{(lm)}_{\{\mu\nu\}}$ with equations of motion \iga. 

For bulk modes we define mass as the value needed to satisfy the on-shell condition
$(-\nabla^2_A + m^2) X=0$ with the understanding that eventually we will go off-shell and consider all eigenvalues of the AdS$_2$ Laplacian $-\nabla^2_A$. This strategy fails for boundary modes since the harmonic equation determines the AdS$_2$ wave function completely from the outset and so the only option will be to go off-shell on $S^2$. We will instead record the spectrum of boundary modes as the eigenvalue of the harmonic operator $(\nabla^2_A+1)C_\mu=m^2C_\mu$ for vectors and 
$(\nabla^2_A+2)H^{(lm)}_{\{\mu\nu\}}=m^2H^{(lm)}_{\{\mu\nu\}}$ for tensors. For boundary modes the ``mass'' becomes a measure of the distance off-shell along $S^2$. With this terminology we find the spectrum

 \vskip .5in

\centering=10em
{
\offinterlineskip
\tabskip=0pt
\halign{ 
\vrule height3.25ex depth1.25ex width 1pt #\tabskip=2em & \hfil#\hfil &\vrule width 1pt  # & \hfil#\hfil &\vrule width 1pt# &  \hfil #\hfil &#\vrule  width 1pt \tabskip=0pt\cr
 \noalign{ \hrule height 1.3pt}
& {\bf Boundary Mode} && Mass && Range & \cr
\noalign{\hrule height 1.0pt}
& $\tilde{B}_{\mu}^{(lm)}$&&   $m^2= l(l+1)+2 $ && $l=1,2\ldots $ &\cr \noalign{\hrule}
& $b^{(lm)}_{\mu}-lB_{\mu}^{(lm)}$ &&   $m^2=l(l-1)  $ && $l=0,1\ldots $ &\cr \noalign{\hrule}
& $b_{\mu}^{(lm)}+(l+1) B^{(lm)}_{\mu}$ &&   $m^2=(l+1)(l+2)  $ && $l=1,2\ldots $ &\cr \noalign{\hrule}
& $H^{(lm)}_{\{\mu\nu\}}$ && $m^2= l(l+1)$ &&$l=0,1\ldots $ &\cr \noalign{\hrule }
&$ \xi^{(lm)}_\mu, c_{\mu}^{(lm)}, \tilde{c}_{\mu}^{(lm)} $ &&  $m^2= l(l+1)+2$  && $l=0,1\ldots $ &\cr \noalign{\hrule}
\noalign{\hrule height 1.3pt}
}}

\vskip .5in

The symmetries of the theory include the tower of 2D diffeomorphisms $\xi^{(lm)}_\mu$. These are 2D vectors so their dualization is also delicate. The residual symmetries remaining after gauge fixing satisfy \heaa, which serves as their equation of motion. We have included these modes in our table along with the ghosts  $c^{(lm)}_\mu$ and anti-ghosts $\tilde{c}^{(lm)}_\mu$ that satisfy the same equations of motion.

We have not yet specified which modes violate the gauge conditions nor have we determined which modes are pure gauge. In the BRST formalism both of these are anyway cancelled by the ghosts and antighosts. The net effect is that the last line in the table (one tower of modes and two ghost towers) cancel the first line in the table (one tower of modes) {\it except} for one mode at $l=0$ that counts with negative sign. The $l=0$ is the spherical reduction of Einstein-Maxwell which is known to have confusing features in AdS$_2\times S^2$. In the present set-up there is $-1$ mode at $l=0$ as one expects from an overconstrained system \ElitzurBF.

We can be more explicit about this. When the 2D diffeomorphisms $\xi^{(lm)}_\mu$ are harmonic they can be dualized to a massless scalar that is not normalizable but such that the vector field itself is normalizable and therefore generates a true symmetry. We can use this symmetry to gauge away the metric components $h_{\mu\alpha}$ with mixed indices on AdS$_2$ and $S^2$. This justifies a physical on-shell approach that simply omits $\tilde{B}_{\mu}^{(lm)}$ and $\xi^{(lm)}_\mu$ from the outset and never introduces ghosts. 

In AdS$_2$ the effective mass is related to conformal weight through $m^2=h(h-1)$. We find that all physical boundary modes have integral conformal weights. 

The dualization of the tensor $H^{(lm)}_{\{\mu\nu\}}$ is less familiar. The harmonic tensors introduced in \ih\ are formally pure gauge generated by a diffeomorphism that can be dualized to a scalar $H_0$ that satisfies $\nabla_A^2 (\nabla^2_A-2)H_0=0$. We can interpret such scalar field as two independent scalars with masses $m^2=0$ and $m^2=2$. The $m^2=0$ component corresponds to non-normalizable scalars that generate a normalizable diffeomorphism. These are precisely the boundary modes that were cancelled two paragraphs ago. On the other hand, the $m^2=2$ component corresponds to non-normalizable scalar modes that generate non-normalizable diffeomorphisms $V_{\mu}$. However, these
non-normalizable diffeomorphisms in turn generate normalizable tensors $H_{\{\mu\nu\}}= \nabla_\mu V_\nu-\nabla_\nu V_\mu - g_{\mu\nu} \nabla^\lambda V_\lambda$. These are physical fields on AdS$_2$ even though they are formally pure gauge. As we discuss in Appendix B, the summation over all modes again produces a volume factor but also a multiplicity factor of three. The tensor thus has {\it three} boundary modes. 

The $m^2=2$ condition on the scalars $H_0$ imply that the non-normalizable vector modes $V_{\mu}$ satisfy 
\eqn\ma{
(\nabla^2_A -1)V_{\mu}=0~.
}
Interestingly, the definition of Conformal Killing Vectors on AdS$_2$ imply this equation. However, the CKVs are precisely those that generate a trivial $H^{(lm)}_{\{\mu\nu\}}$ so the non-normalizable vector modes $V_{\mu}$ are the solutions to \ma\  that are not CKVs on AdS$_2$. 

We introduced the notion of mass for boundary modes as a measure of off-shellness on $S^2$. 
Thus only the $m^2=0$ modes are truly on-shell. In the $b^{(lm)}_{\mu}-lB_{\mu}^{(lm)}$ tower $l=0$ is the mode that is formally pure gauge but with non-normalizable gauge function. For $l=0$ this mode does not mix with gravity and so ``gauge'' really refers to the gauge field and the problem reduces to the spectator vector field discussed in section 2. The $l=1$ mode in the same tower is also massless and again it is formally pure gauge with non-normalizable gauge function. However, the symmetry is a 2D diffeomorphism accompanied by a compensating gauge transformation such that $b^{(1m)}_{\mu}+ 2B_{\mu}^{(1m)}$ is fixed. Specifically this mode is the Conformal Killing Vector $\nabla^\alpha Y_{(1
m)}$ on $S^2$ with a compensating gauge transformation so the gauge field $a^\alpha$ is left invariant. 

The analogous relation between $l=0$ tensors $H^{(00)}_{\{\mu\nu\}}$ and 2D diffeomorphisms was discussed above so all the on-shell boundary modes are related to symmetries. These modes were all previously identified in the discussion of the special cases $l=1$ and $l=0$. We can interpret the full towers of boundary modes as the off-shell realization of these symmetries. This extrapolation to general partial wave number $l$ is nontrivial because of mixing between modes.

\newsec{Quantum Corrections to AdS$_2 \times S^2$ - Bosonic Sector}
Quantum corrections depend only on the spectrum rather than the explicit modes. We consider in turn the contributions from the physical states, the unphysical states, the boundary modes, and the zero modes. We then add the contributions to find the complete heat kernel. 

\subsec{Physical States}
The physical spectrum is
\vskip .5in

\centering=10em
{
\offinterlineskip
\tabskip=0pt
\halign{ 
\vrule height3.25ex depth1.25ex width 1pt #\tabskip=2em & \hfil#\hfil &\vrule width 1pt  # & \hfil#\hfil &\vrule width 1pt# &  \hfil #\hfil &#\vrule  width 1pt \tabskip=0pt\cr
 \noalign{ \hrule height 1.3pt}
&  Mass &&  Multiplicity && Range & \cr
\noalign{\hrule height 1.0pt}
& $m^2=l(l-1)$  &&   $2$ &&$l=2,3\ldots$  &\cr \noalign{\hrule}
& $m^2=(l+1)(l+2)$  &&   $2$ &&$l=1,2\ldots$  &\cr \noalign{\hrule}
\noalign{\hrule height 1.3pt}
}}

\vskip .5in
In each entry the mass refers to the value of $m^2$ such that $(-\nabla^2_A + m^2)X=0$ is the on-shell condition. The bulk result we present agrees with \refs{\MichelsonKN,\LeeYU,\CorleyUZ} \foot{Except that we find the $S^2$ volume mode $\pi^{(00)}$ to be unphysical. This discrepancy was stressed in \KeelerBRA}.
Quantum corrections necessarily consider modes that are off-shell. For modes with $m^2=0$ there is 
a continuous spectrum off-shell with eigenvalues $\lambda\geq {1\over 4}$ for the Euclidean operator $(-\nabla^2_A)$. 
The contributions from this continuous spectrum on AdS$_2$ is encoded in the AdS$_2$ heat kernel \fb. We subsequently sum over the four towers of modes on $S^2$  using \fc. This gives
\eqn\qa{\eqalign{
K^{{\rm bulk},b}_4(s)  &= 2K^s_A(s)\cdot {1\over 4\pi a^2} \cdot \left( \sum_{l=2}^\infty   e^{-sl(l-1)} (2l+1)  
+ \sum_{l=1}^\infty   e^{-s(l+2)(l+1)} (2l+1) \right) \cr
&= K^s_A(s)\cdot {1\over \pi a^2} \left( \sum_{l=0}^\infty   e^{-sl(l+1)} (2l+1) - 1 - {1\over 2}e^{-2s}\right)\cr
& = {1\over 4\pi^2 a^4 s^2} \left( 1 -{3\over 2}s + {137\over 90}s^2 +\ldots \right)
}}

\subsec{Unphysical States}
The full spectrum of modes include some that violate the gauge condition and others that are pure gauge. These two groups of modes coincide precisely. Each has the spectrum 
\vskip .5in

\centering=10em
{
\offinterlineskip
\tabskip=0pt
\halign{ 
\vrule height3.25ex depth1.25ex width 1pt #\tabskip=2em & \hfil#\hfil &\vrule width 1pt  # & \hfil#\hfil &\vrule width 1pt# &  \hfil #\hfil &#\vrule  width 1pt \tabskip=0pt\cr
 \noalign{ \hrule height 1.3pt}
&  Mass &&  Multiplicity && Range & \cr
\noalign{\hrule height 1.0pt}
& $m^2=l(l+1)+2$  &&   $2$ &&$l=0,1\ldots$  &\cr \noalign{\hrule}
& $m^2=l(l+1)$  &&   $1$ &&$l=0,1\ldots$  &\cr \noalign{\hrule}
& $m^2=l(l+1)-2$  &&   $2$ &&$l=1,2\ldots$  &\cr \noalign{\hrule}
\noalign{\hrule height 1.3pt}
}}

\vskip .5in
In our physical quantization scheme we simply omit these modes. They are not allowed even virtually so they do not run in loops. 

In standard covariant quantization we would instead impose the gauge condition and then argue using Ward identities that the pure gauge modes decouple. The upshot will be that indeed these states give no net contribution to the quantum corrections. This structure is of course expected but our construction provides explicit details. 

Similarly, in BRST quantization we allow all the modes and then include $b$ and $c$-ghosts that contribute with opposite sign. These ghost modes will have exactly the same spectrum because they are essentially the pure gauge modes (and their dual constraints). Again there will be no net contribution to the quantum corrections. 

The unphysical modes with $m^2=0$ are special and they are worth discussing. They are the harmonic gauge mode $b^{(00)}_0$, the Conformal Killing Vector on $S^2$ generated by $\zeta^{(1m)}$ and the Killing Vector 
on $S^2$ generated by $\xi^{(1m)}$. Each is a harmonic mode $\nabla^2_A X=0$ on AdS$_2$. The standard covariant quantization above implicitly realizes each of these harmonic modes in both their gradient and curl form. In the off-shell theory these two forms are not equivalent so the two members of the pair are distinct field configurations. Each is equivalent to a massless scalar and the two contributions cancel just as they do for higher $l$. 

The harmonic modes and the Killing Vector on $S^2$ ultimately give boundary states and those we treat differently (in the next subsection). One may therefore object as a matter of principle that the harmonic modes should not be included among the unphysical modes. This question is an ambiguity in the quantization scheme that
does not have a ``correct'' resolution since no physical quantity will depend on it.

\subsec{Boundary Modes}
Each boundary mode receives the constant contribution \bx\ from the AdS$_2$ part. This must be multiplied by
the $S^2$ tower using \fc. The harmonic modes from the two mixed/gravity towers $b^{(lm)}_\mu, B^{(lm)}_\mu$
combine to give
\eqn\qb{\eqalign{
K^{{\rm mix~bndy},b}_4(s)  &= {1\over 2\pi a^2} \cdot {1\over 4\pi a^2} \cdot \left( \sum_{l=0}^\infty   e^{-sl(l-1)} (2l+1)  
+ \sum_{l=1}^\infty   e^{-s(l+2)(l+1)} (2l+1) \right) \cr
&= {1\over 8\pi^2 a^4} \left( 2\sum_{l=0}^\infty   e^{-sl(l+1)} (2l+1) + 2 - e^{-2s}\right)
}}
The harmonic modes from pure gravity reside in the tensors $H_{\{\mu\nu\}}$ (which count with weight three) and in the almost cancelling towers $\tilde{B}^{(lm)}_\mu, \xi^{(lm)}_\mu$. These contributions combine to give
\eqn\qc{\eqalign{
K^{{\rm grav~bndy},b}_4(s)  &= {1\over 2\pi a^2} \cdot {1\over 4\pi a^2} \cdot \left( 3 \sum_{l=0}^\infty + \sum_{l=1}^\infty   e^{-2s}  - \sum_{l=0}^\infty  e^{-2s}   \right)(2l+1) e^{-sl(l+1)}  \cr
& =  {1\over 8\pi^2 a^4} \left( 3 \sum_{l=0}^\infty(2l+1) e^{-sl(l+1)}  - e^{-2s} \right) ~.
}}
The sum of contributions from all bosonic boundary modes becomes 
\eqn\qd{\eqalign{
K^{{\rm bndy},b}_4(s) & =  {1\over 8\pi^2 a^4} \left( 5 \sum_{l=0}^\infty(2l+1) e^{-sl(l+1)} + 2 - 2e^{-2s} \right) \cr
& = {1\over 8\pi^2 a^4 s } \cdot 5(1 + {1\over 3} s + {13\over 15} s^2 +\cdots ) ~.
}}
Ultimately we only need the first two orders. At that precision the boundary modes are equivalent to five free scalar fields on $S^2$. The addition of $2 - 2e^{-2s}$ in the exact result introduces corrections at higher order.

\subsec{Zero modes}
Zero-modes are on-shell boundary modes. They are
\itemitem{$\bullet$}
The pure gauge mode $b^{(00)}_{\mu}$. 

\itemitem{$\bullet$}
The modes $b^{(1m)}_{\mu}-B_{\mu}^{(1m)}$ (with compensating gauge transformation so $b^{(1m)}_{\mu}+2B_{\mu}^{(1m)}$ is fixed) are due to Killing Vectors on $S^2$. These are in the $l=1$ sector so there are $2l+1=3$ modes of this kind.

\itemitem{$\bullet$}
The on-shell modes $H^{(00)}_{\mu\nu}$ are generated by 2D diffeomorphisms on AdS$_2$. The sum over these modes give a multiplicy factor of $3$. 

The zero-modes require special considerations because they are not damped in the Euclidean path integral. As explained in detail by Sen and collaborators, they can be incorporated by a change of variable to the corresponding symmetry parameter \refs{\BanerjeeQC,\SenCJ,\SenDW}. For gauge symmetry it turns out that the naive treatment is correct but for diffeomorphisms the zero modes were undercounted by a factor of two. Each of our $3+3=6$ zero modes that are due to gravity already contributed ${1\over 8\pi^2 a^4}$ but this should be multiplied by two. This correction contributes
\eqn\qe{
K^{{\rm zm},b}_4 = {1\over 8\pi^2 a^4}\cdot 6~,
}
to the heat kernel.

\subsec{Summary}
Adding contributions from bulk (4D), boundary (2D), and the zero-modes (0D) we find
\eqn\qdz{\eqalign{
K^b_4(s) & =  {1\over 4\pi^2 a^4s^2} (1 +  s + {241\over 45} s^2 +\cdots ) ~.
}}
as the total contributions from bosonic modes.

\newsec{Supergravity in AdS$_2 \times S^2$ - Fermionic Sector}

In this section we analyze the two gravitini in ${\cal N}=2$ supergravity in AdS$_{2} \times S^2$. We derive the equations of motion in AdS$_{2}$ point of view via a partial wave expansion and diagonalize them. Only then do we fix the gauge and identify longitudinal states. 


\subsec{4D Theory}

The matter content is a pair of Majorana gravitino fields $\Psi_{IA}$, where $A =1,2$ is an R index. The action for the 4D gravitini is
\eqn\ma{\eqalign{
{\cal L} &= -  \bar{\Psi}_{AI} \Gamma^{IJK}D_J\Psi_{AK}
+{1\over 2}\bar{\Psi}_{AI} \left( F^{IJ}_{AB} + {1\over 2}\Gamma^{IJKL}F_{AB,KL}\right)\Psi_{BJ}~.\cr
}}
We do not bother matching upstairs and downstairs indices when summing over $A,B$. We work with a magnetic background that couples differently to each of the 4D gravitini, so we incorporate index structure in $A,B$: $F_{AB}^{\alpha\beta} = 2\epsilon_{AB}\epsilon^{\alpha\beta}$. 

The supersymmetry that leaves the Lagrangian \ma\ invariant is
\eqn\mb{
\delta\Psi_{AI}  = \left( \delta_{AB} D_I- {1\over 4} {\hat F}_{AB} \gamma_I\right)\theta_B~,
}
for some arbitrary spinor $\theta_B$.

We vary the Lagrangian to obtain the 4D equation of motion,
\eqn\mc{
\Gamma^{IJK} D_J \Psi_{AK} - {1\over 2}( F^{IJ}_{AB} +{1\over 2}\Gamma^{IJKL}F_{AB,KL})\Psi_{JB}=0~.
}

We split the AdS$_2$ and $S^2$-components of the equations of motion, rewrite them in terms of the 2D gamma matrices $\gamma^\mu, \gamma^\alpha$, and use the expression for the background field strength. Our conventions are summarized in Appendix C. The result is
%
%
%
\eqn\md{\eqalign{
\gamma^{\mu\nu}\otimes \gamma^{\alpha}  D_{\nu}   \Psi_{A\alpha}- 
\gamma^{\mu\nu}  \otimes \gamma^{\alpha} D_{\alpha} \Psi_{A\nu}+
\gamma^{\mu} \otimes \gamma^{\alpha\beta} \gamma_S   D_{\alpha} \Psi_{A\beta} +i \gamma^{\mu\nu} \otimes \gamma_S \epsilon_{AB}\Psi_{B\nu}&=0~, \cr
\gamma^{\mu} \otimes \gamma^{\alpha\beta} \gamma_S D_{\beta}  \Psi_{A\mu} - 
\gamma^{\mu} D_{\mu} \otimes \gamma^{\alpha\beta}\gamma_S \Psi_{A\beta}+
\gamma^{\mu\nu}   D_{\mu} \otimes \gamma^{\alpha} \Psi_{A\nu}- \epsilon_{AB}\epsilon^{\alpha\beta}\Psi_{B\beta} &=0~.
}}
Each term is written explicitly as a tensor product to stress that the gamma matrices in AdS and the sphere are in different Clifford algebras and therefore commute. The matrix $\gamma_S$ is the sphere analog of $\Gamma_5$.

\subsec{Partial Wave Expansion.}
We denote spherical spinors with definite angular momentum quantum number $\eta_{(\sigma l m)}$. 
The index $\sigma = \pm$ labels the two components of $\eta_{(\sigma l m)}$. 
A complete set of complex spinors on $S^2$ is then given by $\eta_{(\sigma l m)}$ and $\gamma_S\eta_{(\sigma l m)}$ satisfying \refs{\AbrikosovNJ,\AbrikosovJR}
\eqn\cm{\eqalign{
\gamma^\alpha D_\alpha \eta_{(\sigma l m)} &= i(l+1)\eta_{(\sigma l m)}~,\cr
l&=0,1...
}}
We expand the gravitino wavefunction in spinor spherical harmonics according to
\eqn\me{\eqalign{
\Psi_{A\mu} = \Psi_{+A\mu}^{(\sigma l m)} \otimes \eta_{(\sigma l m)} + \Psi_{-A\mu}^{(\sigma l m)} \otimes \gamma_S \eta_{(\sigma l m)}~,
}}
\eqn\mea{\eqalign{
\Psi_{A\alpha} &= \Psi_{+A}^{(\sigma l m)} \otimes D_{(\alpha)}\eta_{(\sigma l m)} + \Psi_{-A}^{(\sigma l m)} \otimes D_{(\alpha)} \gamma_S \eta_{(\sigma l m)} \cr 
&+ \chi_{+A}^{(\sigma l m)} \otimes \gamma_{\alpha}\eta_{(\sigma l m)} + \chi_{-A}^{(\sigma l m)} \otimes \gamma_{\alpha} \gamma_S \eta_{(\sigma l m)}~.
}}
We expanded the vector index on the gravitino along the sphere in the basis
\eqn\cg{
D_{(\alpha)} \eta_{(\sigma l m)}~,~~~
\gamma_\alpha \eta_{(\sigma l m)}~,
}
where
\eqn\ch{
D_{(\alpha)} = D_\alpha - {1\over 2} \gamma_\alpha \gamma^\beta D_\beta~.
}
The spinors $D_{(\alpha)} \eta_{(\sigma l m)}$ and $\gamma_\alpha \eta_{(\sigma l m)}$ pick out the spin-3/2 part and the spin-1/2 part of the Rarita-Schwinger field on $S^2$. The spin-3/2 part is not defined for $l=0$ so the AdS$_2$ field $\Psi_{\pm A}$ is only defined for $l\geq 1$.

Complex conjugation is given by
\eqn\crb{
\eta_{(\sigma l m) }^*= i \sigma \gamma_S \eta_{(-\sigma l m)}~.
}
The 4D fields $\Psi_{IA}$ are Majorana and thus \crb\ gives the conjugation property
\eqn\cma{
(\Psi_{\pm\mu A}^{(\sigma l m)})^* = \mp i \sigma \Psi_{\mp \mu A}^{(-\sigma l m)}~.
}
The components $\Psi_{\pm A}$ and $\chi_{\pm A}$ transform analogously.

\subsec{Equations of Motion: 2D Theory}
We now insert the spinor harmonic expansion \me\ and \mea\ into the 4D equations of motion \md. We drop the spinor harmonic indices $(\sigma lm)$ to simplify the notation.

We contract the $I =\mu$ equation of motion in \md\ with $\gamma_{\rho\mu}$ then insert the expansion in spinor harmonics. 
\eqn\mg{\eqalign{
0 & = \left (2D_\mu \chi_{-A} + i(l+1)\Psi_{-\mu A} +  {1\over 2}((l+1)^2 -1)\gamma_\mu \Psi_{+A} + i(l+1)\gamma_\mu \chi_{+ A} + i\epsilon_{AB} \Psi_{+\mu B} \right) \otimes \gamma_S \eta \cr
& + \left (2D_\mu \chi_{+A} - i(l+1)\Psi_{+\mu A} +  {1\over 2}((l+1)^2 -1)\gamma_\mu \Psi_{-A} - i(l+1)\gamma_\mu \chi_{- A} + i\epsilon_{AB} \Psi_{-\mu B} \right) \otimes \eta~. 
}}
There is an obvious redundancy in this equation, since the first line is related to the second through complex conjugation. We multiply \mg\ by $(\gamma_S \eta)^\dagger$ and integrate over the sphere coordinates to find
\eqn\mga{\eqalign{
0 & = 2D_\mu \chi_{-A} + i(l+1)\Psi_{-\mu A} +  {1\over 2}((l+1)^2 -1)\gamma_\mu \Psi_{+A} + i(l+1)\gamma_\mu \chi_{+ A} + i\epsilon_{AB} \Psi_{+\mu B}~.
}}
These are the 2D equations of motion. We could alternatively have multiplied by $\eta^\dagger$ and kept the second line of \mg. 

The procedure is repeated for the $I =\alpha$ equations of motion (the second equation in \md). The difference is that the sphere dependent part now carries a vector index. We find
\eqn\mgb{\eqalign{
0& =\left( - \gamma^\mu \Psi_{+A\mu} + \gamma^\mu D_\mu \Psi_{+A} +i\epsilon_{AB}\Psi_{+B} \right) \otimes D^{(\alpha)} \gamma_S \eta \cr
& +\left( - \gamma^\mu \Psi_{-A\mu} + \gamma^\mu D_\mu \Psi_{-A} +i\epsilon_{AB}\Psi_{-B} \right) \otimes D^{(\alpha)}  \eta \cr
& +\left(-{i\over 2}(l+1) \gamma^\mu \Psi_{+A\mu} + \gamma^\mu D_\mu \chi_{+ A} + \gamma^{\mu\nu} D_\mu \Psi_{-A\nu} + i\epsilon_{AB}\chi_{+B}\right)\otimes \gamma^\alpha \gamma_S \eta \cr
& +\left({i\over 2}(l+1) \gamma^\mu \Psi_{-A\mu} + \gamma^\mu D_\mu \chi_{- A} + \gamma^{\mu\nu} D_\mu \Psi_{+A\nu} + i\epsilon_{AB}\chi_{-B}\right)\otimes \gamma^\alpha  \eta ~.
}} 
The operators $D^{(\alpha)}$ and $\gamma^\alpha$ are orthogonal so we can project \mgb\ and integrate over the sphere degrees of freedom,
%
\eqn\mha{\eqalign{
0&=-{i\over 2}(l+1) \gamma^\mu \Psi_{+A\mu} + \gamma^\mu D_\mu \chi_{+ A} + \gamma^{\mu\nu} D_\mu \Psi_{-A\nu} + i\epsilon_{AB}\chi_{+B} 
}}
\eqn\mhb{\eqalign{
0&={1\over 2}[(l+1)^2 -1]\bigg[- \gamma^\mu \Psi_{+A\mu} + \gamma^\mu D_\mu \Psi_{+A} +i\epsilon_{AB}\Psi_{+B} \bigg]
}}
The prefactor $[(l+1)^2 -1]$ in \mhb\ stresses that this equation does not apply for $l=0$. It is analogous to the overall factors of $l(l+1)$ present in some the bosonic sector equations of motion that were not defined at $l=0$.

The complete equations of motion in AdS$_2$ are \mga, \mha, and \mhb. We will work for now with $l = 1,2...$ . The $l=0$ components will be treated separately. 

In order to decouple our equations of motion we define the combinations
\eqn\cra{\eqalign{
\hat{\Psi}_{\mu A} &= \Psi_{+\mu A} -i \Psi_{-\mu A}~, \cr
\hat{\Psi}_{A} &= \Psi_{+ A} -i \Psi_{-A}~, \cr
\hat{\chi}_{A} &= \chi_{+ A} -i \chi_{-A}~,
}}
and the conjugate fields 
\eqn\crbb{\eqalign{
\tilde{\Psi}_{ \mu A} \equiv & \Psi_{+ \mu A} +i \Psi_{- \mu A}~,
}}
with analogous relations defining $\hat{\Psi}_{A}$ and $\hat{\chi}_{A}$. 

Complex conjugation in this basis is given by 
\eqn\crba{\eqalign{
(\hat{\Psi}^{(\sigma l m)}_{\mu A})^* &= -\sigma \tilde{\Psi}^{(-\sigma l m)}_{\mu A}~. \cr
}}
Where we restored the harmonic indices temporarily. The fields $\tilde{\Psi}_{ \mu A}$ are related to $\hat{\Psi}_{\mu A}$ via complex conjugation according to \crba. The fields $\tilde{\Psi}_{ \mu A}$ present no new information.

By inspection of the equations of motion we see that the 2D Rarita-Schwinger field $\Psi_{\mu A}$ is dependent on the fields $\Psi_A$ and $\chi_A$. Hence, we use \mga\ to express $\Psi_{\mu A}$ in terms of the other modes and simplify the remaining equations \mha\ and \mhb.

Recall that the index $A$ takes two values, and for each field such as  $\Psi_{+ \mu A}$ there is a complex conjugate $\Psi_{- \mu A}$. Thus, we are looking into four vector valued equations. It is somewhat tedious yet straight forward to write all four equations in components then solve for each $\Psi_{\pm \mu  A}$. The result in the basis \cra\ is
\eqn\mj{
\hat{\Psi}_{\mu A} = {-i \over 1 - (l+1)^2}(-i(l+1)\delta_{AB} + \epsilon_{AB}) \bigg(-2i D_\mu \tilde{\chi}_B - {1-(l+1)^2\over 2}\gamma_\mu \hat{\Psi}_B + i(l+1)\gamma_\mu \tilde{\chi}_B\bigg)~,
}
and similarly for the conjugate field $\tilde{\Psi}_{\mu A}$. We will refer to \mj\ as the Rarita-Schwinger 
constraint. Note that it cannot be continued to $l=0$ which we study separately. 

We now insert the Rarita-Schwinger constraint \mj\ into the equations of motion \mha\ and \mhb. The first order derivative in \mj\ is acted on by further derivatives but the resulting second order term appears as a commutator that reduces to a curvature factor. The resulting equations are therefore of first order:
\eqn\mk{\eqalign{
\bigg(\gamma^\mu D_\mu - (l+1)\bigg) \bigg[ \hat{\Psi}_{A}  + {2\over(l+1)^2 -1}(i(l+1)\delta_{AB} - \epsilon_{AB} )\tilde{\chi}_{B} \bigg] =0~,
}}
\eqn\mka{\eqalign{
\bigg(\gamma^\mu D_\mu - (l+1)\bigg) (i(l+1)\delta_{AB} - \epsilon_{AB} ) \bigg[ 2\hat{\Psi}_{B}  + {1\over(l+1)^2 -1}(i(l+1)\delta_{BC} - \epsilon_{BC} )\tilde{\chi}_{C} \bigg] =0~.
}}
The operator $(i(l+1)\delta_{AB} - \epsilon_{AB} )$ can be inverted for $l\neq0$, so we can decouple these into Dirac equations for $\hat{\Psi}_{A}$ and $\tilde{\chi}_{A}$:
\eqn\ml{\eqalign{
&(\gamma^\mu D_\mu - (l+1)) \hat{\Psi}_{A}  =0~,\cr
&(\gamma^\mu D_\mu - (l+1)) \tilde{\chi}_{A}  =0~.
}}
The conjugate equations similarly give
\eqn\mla{\eqalign{
&(\gamma^\mu D_\mu + (l+1)) \tilde{\Psi}_{A}  =0~,\cr
&(\gamma^\mu D_\mu + (l+1)) \hat{\chi}_{A}  =0~.
}}
At this point we have successfully decoupled all equations of motion with no constraints or gauge 
condition imposed.

\subsec{Dualization.}

We showed above that the field $\hat{\Psi}_{\mu A}$ is not independent from the spinors $\hat{\Psi}_{A}$ and $\tilde{\chi}_{A}$. However, we are going to fix a gauge and study supersymmetry variations that involve components of $\hat{\Psi}_{\mu A}$. So instead of throwing away the vector-spinors $\hat{\Psi}_{\mu A}$ we will dualize them into spinors in order to more precisely work with the Rarita-Schwinger constraint \mj, gauge conditions, and variations.

We dualize $\hat{\Psi}_{\mu A}$ according to
\eqn\mm{
\hat{\Psi}_{\mu A} = D_{(\mu)} \hat{\kappa}_A + \gamma_\mu \hat{\tau}_A~.
}
Where $D_{(\mu)} = D_\mu - {1\over2}\gamma_\mu \gamma^\nu D_\nu$. An analogous dualization is carried for $\tilde{\Psi}_{\mu A}$. Our field content is then the 16 components: $\hat{\kappa}_A, \hat{\tau}_A, \hat{\Psi}_A, \hat{\chi}_A,$ with $A =1,2$ and their conjugates with tildes. 

%
%
We can recast the Rarita-Schwinger constraint \mj\ as equations expressing the dual spinors introduced in \mm\ to other field components:
\eqn\mmb{\eqalign{
\hat{\kappa}_A &= {2 \over 1-(l+1)^2} (i(l+1)\delta_{AB} - \epsilon_{AB})\tilde{\chi}_{B}~, \cr
\hat{\tau}_A &= -{i\over2}(i(l+1)\delta_{AB} - \epsilon_{AB})\hat{\Psi}_B~, \cr
\tilde{\kappa}_A &= {2 \over 1-(l+1)^2} (i(l+1)\delta_{AB} + \epsilon_{AB})\hat{\chi}_{B}~, \cr
\tilde{\tau}_A &={i\over2}(i(l+1)\delta_{AB} + \epsilon_{AB})\tilde{\Psi}_B ~.
}}
This is the dual form of the result that we can eliminate half of the initial field components and only work with the components $\hat{\Psi}_A, \tilde{\Psi}_A, \hat{\chi}_A, \tilde{\chi}_A$. This formulation will be useful in the following section.


\subsec{Gauge Violating, Longitudinal, and Physical States.}

We now impose Lorentz gauge on the on shell states we found and then construct pure gauge states. 

The Lorentz gauge condition is $\Gamma^I \Psi_I =0$. We write it in terms of 2D gamma matrices, insert the expansion of $\Psi_I$ in spherical spinors, and dualize according to \mm. The gauge condition in terms of 2D spinors is
\eqn\mn{\eqalign{
\hat{\tau}_A -i \tilde{\chi}_A &=0~,\cr
\tilde{\tau}_A +i \hat{\chi}_A &=0~.\cr
}}
We already have expressed $\hat{\tau}_A$ and $\tilde{\tau}_A$ in terms of other fields in \mmb\ so we can write the gauge condition in terms of $\hat{\Psi}_A, \tilde{\chi}_A$ and their conjugates
\eqn\mna{\eqalign{
\tilde{\chi}_A &=  -{1\over2}(i(l+1)\delta_{AB} - \epsilon_{AB})\hat{\Psi}_B ~,\cr
\hat{\chi}_{A} &=-{1\over 2}(i(l+1)\delta_{AB} + \epsilon_{AB})\tilde{\Psi}_B  ~.\cr
}}
After imposing the equations of motion and gauge condition there are four field components: $\hat{\Psi}_A, \tilde{\Psi}_A$, $A=1,2$.

 We now look for pure gauge states. The supersymmetry variations of the 4D Rarita-Schwinger fields $\Psi_{IA}$ are given by
\eqn\mnb{\eqalign{
\delta \Psi_{IA} &= (D_I \delta_{AB} - {1\over 4} \Gamma_{JK}F^{JK}_{AB} \Gamma_I )\theta_B\cr
&= (D_I \delta_{AB} + {i\over 2} (1\otimes \gamma_S) \Gamma_I \epsilon_{AB})\theta_B~.
}}
In order to compute the supersymmetric variation of each mode we expand the spinor $\theta_A$ into partial waves in analogy with \me\ - \mea,
\eqn\mnc{
\theta_A = \theta_{+A} \otimes \eta +  \theta_{-A} \otimes \gamma_S \eta~,
}
and rewrite the $\pm$ indices as the combinations $\hat{\theta}_A$ and $\tilde{\theta}_A$:
\eqn\mnd{\eqalign{
\hat{\theta}_{A} &= \theta_{+ A} -i \theta_{-A}~, \cr
\tilde{\theta}_{A} &= \theta_{+ A} +i \theta_{-A}~.
}}
Note that the procedure here is in complete analogy with the bosonic sector: one writes the gauge variations then expands the parameters in partial waves. The next step is to find the constraints the gauge condition imposes on the supersymmetric parameters, that is, the residual gauge symmetry. 

The preservation of the Lorentz gauge condition $\Gamma_I\Psi^I=0$ constrains the 
4D supersymmetric parameters to satisfy
\eqn\mne{
\Gamma^I\big[D_I \delta_{AB} + {i\over 2} (1\otimes \gamma_S) \Gamma_I \epsilon_{AB}\big]\theta_B = 0~.
}
Expression \mne\ is once again decomposed into 2D conditions. The result are the constraints
\eqn\mnf{\eqalign{
( \gamma^\mu D_\mu + (l+1) )\hat{\theta}_A &=0~, \cr
( \gamma^\mu D_\mu - (l+1) )\tilde{\theta}_A &=0~.
}}
The residual gauge symmetry has to satisfy \mnf\ in order not to violate the imposed gauge. 

We compute the supersymmetric variations of the dualized spinors in terms of the parameters $\hat{\theta}_{A} $, $\tilde{\theta}_{A}$, by expanding both sides of \mnb\ in spinor harmonics, dualizing when needed, and comparing each variation in the \mnd\ basis. We get
\eqn\mng{\eqalign{
\delta \hat{\kappa}_A &= \hat{\theta}_A~, \cr
\delta \tilde{\kappa}_A &=\tilde{\theta}_A~, \cr
\delta  \hat{\tau}_A &= {1\over 2} (\gamma^\mu D_\mu \delta_{AB} +i\epsilon_{AB})\hat{\theta}_B~, \cr
\delta  \tilde{\tau}_A &= {1\over 2} (\gamma^\mu D_\mu \delta_{AB} +i\epsilon_{AB})\tilde{\theta}_B~, \cr
\delta  \hat{\Psi}_A &= \hat{\theta}_A~, \cr
\delta \tilde{\Psi}_A &= \tilde{\theta}_A~, \cr
\delta \hat{\chi}_A &= {1\over 2} (i(l+1) \delta_{AB} -\epsilon_{AB})\tilde{\theta}_B~, \cr
 \delta \tilde{\chi}_A &= {1\over 2} (i(l+1) \delta_{AB} +\epsilon_{AB})\hat{\theta}_B~.
}}
We cannot remove $\hat{\Psi}_{A}$ and $\tilde{\Psi}_{A}$ with residual gauge transformations since their equations of motion \ml- \mla\ are inconsistent with \mnf. 

As a clarifying example consider the 4D flat space case: supersymmetry transformations are given by $\delta \Psi_I = \partial_I \theta $ and the gauge condition $\gamma^I \Psi_I=0$ requires $\theta$ to be massless. One cannot turn on pure gauge modes with a massive parameter $\theta$ since those would be gauge violating. An analogous situation is happening here. We cannot gauge away modes using the residual symmetry we have. Thus, there are no longitudinal modes.

The modes $\hat{\Psi}_{A}$, $\tilde{\Psi}_{A}$ with $A=1,2$, $l\geq 1$, and the masses reported in \ml, \mla\ satisfy the gauge condition and are not gauge equivalent to vacuum. They are physical modes. This result agrees with \refs{\CorleyUZ,\KeelerBRA}.

\subsec{l=0 Modes.}
In this section we analyze the $l=0$ sector. Two related issues that are special to $l=0$ change the equations that apply: the $\Psi_{\pm A}$ components of the gravitino are not defined and also the equation of motion \mhb\ does not apply. We are therefore left with \mga\ and \mha\ which we write in the ``hat-tilde" basis as
\eqn\mpb{-{i\over 2} \gamma^\mu \hat{\Psi}_{A\mu} + \gamma^\mu D_\mu \tilde{\chi}_{ A} + i \gamma^{\mu\nu} D_\mu \hat{\Psi}_{A\nu} + i\epsilon_{AB}\tilde{\chi}_{B} =0~.
}
\eqn\mpa{
\bigg( D_\mu - {1\over 2}\gamma_\mu \bigg) \tilde{\chi}_{A} = {1\over 2}(i \delta_{AB} + \epsilon_{AB})\hat{\Psi}_{\mu B}~. 
}
There are also analogous expressions for the conjugate field. Contracting these equations with the projection operators $(i\delta_{AB} \pm \epsilon_{AB})$ we find
\eqn\mqc{
(i\delta_{AB} + \epsilon_{AB})\bigg[-{i \over 2} \gamma^\mu \hat{\Psi}_{B\mu} +i \gamma^{\mu\nu}D_\mu \hat{\Psi}_{B\nu}  + (\gamma^\mu D_\mu -1)\tilde{\chi}_B\bigg] = 0~, 
}
\eqn\mqe{(i\delta_{AB} + \epsilon_{AB})\bigg[ (D_{\mu}-{1\over 2}\gamma_\mu)\tilde{\chi}_B - i \hat{\Psi}_{B\mu} \bigg] =0~.
}
\eqn\mqf{
(i\delta_{AB} - \epsilon_{AB})\bigg[- {i \over 2} \gamma^\mu \hat{\Psi}_{B\mu} +i \gamma^{\mu\nu}D_\mu \hat{\Psi}_{B\nu}  + (\gamma^\mu D_\mu +1)\tilde{\chi}_B\bigg] = 0~, 
}
 \eqn\mqh{(i\delta_{AB} - \epsilon_{AB}) (D_{\mu}-{1\over 2}\gamma_\mu)\tilde{\chi}_B =0 ~.
}

We next impose Lorentz gauge in the form
\eqn\mqbe{
 \gamma^\mu \hat{\Psi}_{A\mu} = 2i\tilde{\chi}_A~.
}
The gauge fixed gravitino equations then simplify to 

\eqn\mqi{
(i\delta_{AB} \pm \epsilon_{AB})( D^{\mu} +{1\over 2} \gamma^\mu) \hat{\Psi}_{A\mu} = 0~.
}
We still have the equations of motion \mqe\ and \mqh\ for $\tilde{\chi}_A$. 

In the sector with $(i\delta_{AB} +\epsilon_{AB})$ projection the equation of motion \mqe\ and the gauge condition \mqbe\ combine to give
\eqn\mqr{
(i\delta_{AB} +\epsilon_{AB}) (\gamma^\mu D_\mu + 1 ) \tilde{\chi}_A = 0~.
}
Given a solution to this equation we can specify the gravitino $\hat{\Psi}_{A\mu}$ as in \mqe\ and then the gauge condition and the gravitino equation \mqi\ are all satisfied. Thus solutions to \mqr\ parametrize the space of solutions to the full equations. It can be shown that all these solutions are pure gauge (up to normalization issues). We stress for later that in the special case where $\tilde{\chi}_A$ vanishes the gravitino $\hat{\Psi}_{A\mu}$ vanishes as well. 

The sector with $(i\delta_{AB} -\epsilon_{AB})$ projection is more involved. Here \mqh\ specifies $\tilde{\chi}_A$ as a Killing Spinor in AdS$_2$ with mass $+1$:
\eqn\mqrc{
(i\delta_{AB} -\epsilon_{AB}) (\gamma^\mu D_\mu - 1 ) \tilde{\chi}_A = 0~.
}
The gauge condition \mqbe\ (which we could represent in terms of dual fields as in \mn) then gives the trace part of the gravitino but the traceless part remains unspecified. Rewriting the gravitino equation of motion \mqi\  in terms of the dual spinor $\hat{\kappa}_A$ introduced in \mm\ we have
\eqn\mqra{
(i\delta_{AB} -\epsilon_{AB}) \left( [(\gamma^\mu D_\mu )^2-1]\hat{\kappa}_A - 4i \tilde{\chi}_A\right)=0~.
}
Given the Killing spinor $\tilde{\chi}_A$ this equation permits a particular solution for $\hat{\kappa}_A$. To this solution we can add solutions to the homogenous equation which we can represent as solutions to 
\eqn\mqrb{
(i\delta_{AB} -\epsilon_{AB})(\gamma^\mu D_\mu \pm 1 )\hat{\kappa}_A=0~,
}
with either sign. In the special case where $\tilde{\chi}_A$ vanishes the traceless component of the gravitino is given by solutions to these equations. 

The lightest fermion masses $\pm 1$ are special in that they correspond to zero modes of the Dirac operator squared. The Euclidean version of these modes do not comprise a continuum of solutions of plane wave type but rather a discrete set of modes which are necessarily nonnormalizable. For this reason only the solutions with $\tilde{\chi}_A=0$ are physical. After this normalizability condition is imposed the space of $l=0$ modes that satisfy the equations of motion and the gauge condition reduces to the solutions of \mqrb. Although these fields are also non-normalizable they are dual to physical gravitini 
\eqn\mqrc{
(i\delta_{AB} - \epsilon_{AB}) \hat{\Psi}_{A\mu}=(i\delta_{AB} - \epsilon_{AB}) D_{(\mu)} \hat{\kappa}_A 
= (i\delta_{AB} - \epsilon_{AB}) (D_\mu \pm {1\over 2} \gamma_\mu ) \hat{\kappa}_A~,
}
that are normalizable in addition to satisfying the equation of motion and the gauge condition. The  
$\hat{\kappa}_A$ is such that $\gamma^\mu\hat{\Psi}_{A\mu}=0$.

We finally need to ask whether the remaining modes \mqrc\ are longitudinal. The pure gauge modes are 
\eqn\mqp{
(i\delta_{AB} - \epsilon_{AB}) \delta \hat{\Psi}_{B\mu} = (i\delta_{AB} - \epsilon_{AB})  (D_{\mu} +{1\over 2} \gamma_\mu)\hat{\theta}_B~.
}
with the residual SUSY transformation such that it preserves the gauge condition 
\eqn\mqq{
(\gamma^\mu D_{\mu} +1)\hat{\theta}_A =0~. 
}
The mode that appears with upper sign in \mqrb\ is therefore pure gauge with the field and the gauge parameter coinciding
$\hat{\kappa}_A=\hat{\theta}_A$ as we expected from \mng. Since the gauge parameter is not normalizable the corresponding gravitino is physical even though it is formally pure gauge. 

The mode that appears with lower sign in \mqrc\ is similarly nonnormalizable but corresponding to a normalizable gravitino. This mode is again formally pure gauge but with a transformation parameter that does not satisfy the condition \mqq\ that the gauge is preserved. It is therefore not pure gauge because the would-be gauge transformation introduces a nonvanishing $\gamma^\mu\hat{\Psi}_{A\mu}$. It is possible to instead define a superconformal symmetry that leaves $\gamma^\mu\hat{\Psi}_{A\mu}$ invariant and consider this mode pure gauge with respect to this extended symmetry. Either way, it is a physical boundary mode. 

Recall that the computation in this subsection focussed for definiteness on the $\hat{\Psi}_{A\mu}$, $\tilde{\chi}_{A}$ field components. It can be repeated for the conjugate fields $\tilde{\Psi}_{A\mu}$. $\hat{\chi}_{A}$. The analogue of \mqrc\ in this sector is
\eqn\mqrt{
(i\delta_{AB} + \epsilon_{AB}) \tilde{\Psi}_{A\mu}=(i\delta_{AB} + \epsilon_{AB}) D_{(\mu)} \tilde{\kappa}_A 
= (i\delta_{AB} + \epsilon_{AB}) (D_\mu \pm {1\over 2} \gamma_\mu ) \tilde{\kappa}_A~,
}
with $\tilde{\kappa}_A$ such that $\gamma^\mu\tilde{\Psi}_{A\mu}=0$. It is the opposite SUSY that gives rise to a boundary mode and it is now the lower sign that is a pure gauge mode while the upper is a superconformal extension. 

In summary, there are no physical bulk modes at $l=0$. However, each of the two SUSYs allow a nonnormalizable gauge parameter (and a superconformal analogue) that generates normalizable gravitini. This corresponds to four physical boundary modes. 

A more detailed discussion on the normalizability  of fermionic boundary modes is found at Appendix C.

\newsec{Quantum Corrections to AdS$_2 \times S^2$ - Fermionic Sector}

In this section we compute the heat kernels for the fermionic sector of the gravity multiplet. An important preliminary result is the heat kernel of a free spin 1/2 fermion on the sphere $S^2$,
\eqn\muaa{
K^f_S = {1\over 4\pi a^2}\sum_{k=0}^\infty e^{-s(k+1)^2}(2k+2) =  {1\over 4\pi a^2s}\bigg( 1 -{1\over6}s -{1\over60}s^2 +...\bigg)~.
}
The AdS$_{2}$ heat kernel is obtained to the precision we need by flipping the sign of the terms that are odd in the curvature (with the overall sign changed due to fermion statistics)
\eqn\mua{
K^f_A = -{1\over 4\pi a^2}\sum_{k=0}^\infty e^{-s(k+1)^2}(2k+2) =  -{1\over 4\pi a^2s}\bigg( 1 +{1\over6}s -{1\over60}s^2 +...\bigg)~.
}
As in \fd\ for bosons we compute the 4D heat kernels by summing over towers using
\eqn\mub{
K^f_4 = K^f_A \cdot {1\over 4\pi a^2} \sum_j e^{-m^2_j s}(2j+2)~.
}
We are summing over each value of the angular momentum $j$ on $S^2$ weighed by the effective AdS$_{2}$ masses.

\subsec{Physical States}

The physical bulk spectrum summarized at the end of section 6.5 is four fermionic bulk degrees of freedom with masses $m^2 = (k+1)^2$ where $k>0$. Hence, the 4D heat kernel is
\eqn\muc{\eqalign{
K^{\rm bulk}_4 &= 4 \cdot K^f_A \cdot  {1\over 4\pi a^2} \sum_{k=1}^\infty e^{-s(k+1)^2}(2k+2) \cr
&= 4 \cdot K^f_A \cdot  {1\over 4\pi a^2} \bigg(\sum_{k=0}^\infty e^{-s(k+1)^2}(2k+2) -2 e^{-s}  \bigg) \cr
& = - {1 \over 4\pi^2 a^4 s^2} \bigg( 1 - {11\over180 }s^2 + ... - 2s \big( 1 - {5\over6}s \big)+ ....\bigg)~. 
}}
We wrote the final line as the sum of the result we would get from four free fermionic degrees of freedom and a term we interpret as due to the couplings of the gravitino field.

\subsec{Unphysical States}

The unphysical spectrum consists of twelve fermionic bulk degrees of freedom with masses $m^2 = (k+1)^2$ at $k\geq0$. These modes were all established as unphysical either due to the Rarita-Schwinger constraint -- which is a component of the equations of motion --  or due to the gauge condition. No on-shell modes were removed by residual gauge symmetries. In our on-shell method we do not include contributions from any of these. 

\subsec{Boundary Modes}

The boundary modes are zero modes in AdS$_2$ while consisting of a full tower on $S^2$. Expression \mub\ for a 4D heat kernel is then modified to
\eqn\mud{
K^{\rm bndy}_4 =  -{1\over 8 \pi^2 a^4} \sum_j e^{-m^2_j s}(2j+2)~,
}
where the contribution of the AdS$_2$ heat kernel is a factor of the regulated volume of AdS. 

The boundary fields $\hat{\theta}_A, \tilde{\theta}_A$ each have a projection on the $R$ index $A$ but also a doubling due to conformal symmetry. Thus there are four towers of boundary states. We used the mass \mnf\ to find the mass squared and then the heat kernel
\eqn\mue{\eqalign{
(\gamma^\mu D_\mu)^2 D_{(\nu)}\hat{\theta}_A &= D_{(\nu)} [ (\gamma^\mu D_\mu)^2 -1] \hat{\theta}_A \cr
& =  [ (k+1)^2 -1] D_{(\nu)} \hat{\theta}_A~.
}}
The total heat kernel for the four boundary modes then becomes
\eqn\muf{\eqalign{
K^{\rm bndy}_4 &=  -{4 \over 8 \pi^2 a^4}  \sum_{k=0}^\infty e^{- [ (k+1)^2 -1]s}(2k+2)\cr
& = -{4 \over 8 \pi^2 a^4} \bigg({1\over s} - {1\over 6}\bigg) e^{-s} \cr
& = -{1 \over 4 \pi^2 a^4} \bigg({2\over s} + {5\over 3} + \ldots\bigg) ~.
}}

\subsec{Zero Modes}

Boundary states that are also zero modes on the $S^2$ are true zero modes of AdS$_2 \times S^2$.  Hence, the zero mode content can be read off from the spectrum of boundary states. The four fermionic zero-modes are the $k=0$ entries in \mue. As mentioned in the bosonic sector, zero-modes require special considerations discussed by \refs{\BanerjeeQC,\SenCJ,\SenDW}.

In the naive treatment \muf\ each of the four zero modes contributes with $- {2\over 8\pi^2a^2}$, but the correct contribution is larger. The correction due to zero-modes is 
\eqn\mug{
K^{zm, f}_4 =  - {8\over  8\pi^2 a^4}  \bigg({3\over2}- {1\over2}\bigg) 
= {1\over  8\pi^2 a^4}  \cdot ( -8 )~.
}

\subsec{Summary.}
We add the fermionic contributions from bulk (4D), boundary (2D), and the zero-modes (0D),
\eqn\muh{
K^f_{4} = - {1 \over 4\pi^2 a^4 s^2} \bigg( 1 +{1309\over180 }s^2 + \ldots\bigg)~,
}
which is the total contribution from fermionic modes.

We finally add the total bosonic contribution \qdz\ and the total fermionic contribution \muh,

\eqn\mui{\eqalign{
K^b_4 + K^f_{4} =   {1\over 4\pi^2 a^4} \bigg(  {1\over s} - {23\over 12} +\ldots \bigg)~.
}}
These are the quantum corrections to supergravity in AdS$_2\times S^2$. 


\medskip

\noindent
{\bf Acknowledgement}

We thank A. Charles, C. Keeler, M. Marino, A. Sen, M. Spradlin, A. Strominger for discussions. This work was supported in by the US Department of Energy under grant DE-FG02-95ER40899.

\newsec{Appendix A: Generalized Eigenvectors.}

Repeated eigenvalues and generalized eigenvectors play an important role in our solutions so here we review a few of their features. 

An elementary example with an eigenvalue that is repeated twice is the nonhermitean $2\times 2$ matrix
\eqn\zb{ 
M = \pmatrix{2 & 1 \cr 0 & 2}~,
}
with two eigenvalues identical to 2. There is only one true eigenvector
\eqn\zbc{
\eta_1 = \pmatrix{1 \cr 0}~, 
}
but there also a generalized eigenvector 
\eqn\zbd{
\eta_2 = \pmatrix{0\cr 1}~,
}
that satisfies the generalized eigenvalue equation
\eqn\zc{(M - \lambda I_2 )^2\eta_2 =0~,
}
with eigenvalue $\lambda=2$. The generalized eigenvector $\eta_2$ is not a true eigenvector since
\eqn\zca{
(M - \lambda I_2 ) \eta_2 =  \pmatrix{0 & 1 \cr 0 & 0}\eta_2  = \eta_1~. 
}
However, the generalized eigenvalue equation \zc\ follows because $\eta_1$ is a true eigenvector. Importantly, the determinant ${\rm det~M} = 2\cdot 2=4$ is the product of eigenvalues even though one appearance of the repeated eigenvalue $\lambda=2$ only allows a generalized eigenvector. 

Generalized eigenvectors are ubiquitous in our setting because the linearized equations of motion have kinetic terms and mass-matrices that cannot be simultaneously diagonalized. For example, the AdS$_2$ volume mode 
$H_{\phantom{(00)}\rho}^{(00)\phantom{\rho}\rho}$ and the $S^2$ volume mode $h_\alpha^{~\alpha} = \pi^{(00)}$ couple through the Lagrangean
\eqn\zcb{
 {\cal L}_{\rm scalar}^{l=0} =  -{1\over 8}H_{\phantom{(00)}\rho}^{(00)\phantom{\rho}\rho} (\nabla^2_A - 2) \pi^{(00)} - {1\over 4}\pi^{(00)2} ~.
 }
In the given basis the mass matrix is diagonal but the kinetic matrix is not. There is no basis where both are diagonal. The equations of motion are naturally presented in a form where  $\pi^{(00)}$ sources $H_{\phantom{(00)}\rho}^{(00)\phantom{\rho}\rho}$ but not the other way around
\eqn\zce{
\nabla^2_x \pmatrix{ H_{\phantom{(00)}\rho}^{(00)\phantom{\rho}\rho} \cr  \pi^{(00)}} = \pmatrix{ 2 & -4 \cr 0 & 2 }\pmatrix{ H_{\phantom{(00)}\rho}^{(00)\phantom{\rho}\rho} \cr  \pi^{(00)}}~.
}
The mass matrix is similar to \zb\ and the eigenvalue problem is analogous to the elementary one discussed above. $\pi^{(00)}$ is a true eigenvector but $H_{\phantom{(00)}\rho}^{(00)\phantom{\rho}\rho}$ is just a generalized eigenvector satisfying 
 \eqn\zcg{\eqalign{
 &(\nabla^2_x -2)^2H_{\phantom{(00)}\rho}^{(00)\phantom{\rho}\rho}  =0~.
}}

We consider one additional example from our setting: the fields $b_{\parallel}^{(lm)}$, $B^{(lm)}_{\parallel}$, $\tilde{b}^{(lm)}$ for $l\geq 1$. The equations of motion \ihah:
\eqn\zfa{
(\nabla^2_x - l(l+1))\pmatrix{B_{\parallel}^{(lm)}  \cr  b^{(lm)}_{\parallel} \cr \tilde{b}^{(lm)}} = \pmatrix{ 2 & 2 & -2 \cr 2l(l+1)& 0 & 0 \cr 4+2l(l+1) & 4 & -4 }\pmatrix{ B_{\parallel}^{(lm)}  \cr  b^{(lm)}_{\parallel} \cr \tilde{b}^{(lm)}}~.
}
The $3\times 3$ matrix on the RHS of \zfa\ has one eigenvalue $\lambda=-2$ and also a repeated eigenvalue $\lambda=0$. There are two conventional (true) eigenvectors and one generalized eigenvector:  
\vskip .5in

\centering=10em
{
\offinterlineskip
\tabskip=0pt
\halign{ 
\vrule height3.25ex depth1.25ex width 1pt #\tabskip=2em & \hfil#\hfil &\vrule width 1pt  # & \hfil#\hfil &\vrule width 1pt# &  \hfil #\hfil &#\vrule  width 1pt \tabskip=0pt\cr
 \noalign{ \hrule height 1.3pt}
& {\bf Mode} && Mass  && Comment & \cr
\noalign{\hrule height 1.0pt}
& $b_{\parallel}^{(lm)} + B^{(lm)}_{\parallel} - \tilde{b}^{(lm)} $  && $m^2= l(l+1)-2$   && Conventional.&\cr \noalign{\hrule}
& $ b_{\parallel}^{(lm)} + 2B^{(lm)}_{\parallel}- \tilde{b}^{(lm)} $  && $m^2= l(l+1)$ && Conventional.&\cr \noalign{\hrule}
& $b_{\parallel}^{(lm)}  +l(l+1) B^{(lm)}_{\parallel}$  && $m^2= l(l+1)$ &&   Generalized.&\cr \noalign{\hrule}
\noalign{\hrule height 1.3pt}
}}

\vskip .5in

The generalized eigenvector satisfies 
\eqn\zeb{\eqalign{
[\nabla^2_A -l(l+1)](b_{\parallel}^{(lm)}  +l(l+1) B^{(lm)}_{\parallel}) = - (b_{\parallel}^{(lm)} +2B^{(lm)}_{\parallel}-\tilde{b}^{(lm)}) ~.
}}
The RHS is a true eigenvector of $[\nabla^2_A -l(l+1)]$ with eigenvalue $\lambda=0$ so the higher order operator 
$[\nabla^2_A -l(l+1)]^2$ annihilates the generalized eigenvector $b_{\parallel}^{(lm)}  +l(l+1) B^{(lm)}_{\parallel}$.
 
The contribution to the functional determinant from these fields is computed correctly by multiplication of all eigenvalues whether they are repeated or not. Thus, the complications due to generalized eigenvectors are not an issue as far as the heat kernels are concerned.

\newsec{Appendix B: Tensor Modes on the Boundary}
We want to identify residual diffeomorphisms that are not fixed by our gauge. A 2D diffeomorphism generated by $\xi_\mu$ gives rise to a traceless symmetric tensor 
\eqn\wa{
H_{\{\mu\nu\}} = \nabla_\mu\xi_\nu + \nabla_\mu\xi_\nu - g_{\mu\nu} \nabla_\rho \xi^\rho~.
}
The gauge condition $\nabla^\mu H_{\{\mu\nu\}}={1\over 2} \nabla_\nu \pi$ with the 2D scalar $\pi$ invariant is preserved  iff the vector $\xi_\mu$ satisfies 
\eqn\wb{
(\nabla^2_A  - 1)\xi_\mu=0~.
}
For K\"{a}hler metrics on the disc we can rewrite the holomorphic component of \wb\ as
\eqn\wc{
2g^{z{\bar z}}  \nabla_{\bar z} \nabla_z\xi _z=0~.
}
The covariant derivative is $\nabla_{\bar z}=\partial_{\bar z}$ when acting on an object with lower holomorphic indices so the 
solutions are those where $\nabla_z\xi _z$ are holomorphic. The induced tensor $H_{\{\mu\nu\}}$ is therefore a {\it quadratic holomorphic differential}. 

We consider the holomorphic differential $\nabla_z\xi_z =  z^{n-2}$ with $n\geq 2$. The holomorphic derivative is 
\eqn\wd{
\nabla_z\xi _z = g_{z{\bar z}} \partial_z ( g^{z{\bar z}} \xi_z)= g_{z{\bar z}} \partial_z \xi^{\bar z}~,
}
so 
\eqn\we{
\partial_z \xi^{\bar z} = {1\over 2a} ( 1 - |z|^2)^2 z^{n-2}~,
}
and upon integration we find
\eqn\wf{
 \xi^{\bar z} = {1\over 2a} \left( {1\over n-1} z^{n-1} - {2{\bar z}\over n} z^n + {{\bar z}^2\over n+1} z^{n+1}\right)~. 
}
This explicit form shows that we must indeed take $n\neq 0,\pm 1$~. For $n\geq 2$ the vector exists but it is not normalizable 
\eqn\wh{
\int_{|z|\leq 1} |\xi_z|^2 \sqrt{g} d^2 z = \int_{|z|\leq 1} | \xi_z \xi_z^*| ~d^2 z = \int_{|z|\leq 1} ~|g_{z{\bar z}}  \xi^{\bar z} |^2 ~d^2 z \to \infty~,
}
since $g_{z{\bar z}}$ diverges as $|z|\to 1$ while $|\xi^z|$ remains finite. 

Importantly the quadratic holomorphic differential generated by the non-normalizable vector is finite 
\eqn\wg{
\int_{|z|\leq 1} |\nabla_z \xi_z|^2 ~ \sqrt{g} d^2 z = \int_{|z|\leq 1} g^{z{\bar z}} |z|^{2(n-2)} ~d^2 z < \infty ~,
}
for $n\geq 2$ since $g^{z{\bar z}} = {1\over 2a} ( 1 - |z|^2)^2$ is perfectly well behaved near the boundary at $|z|=1$. We introduce the tensor modes 
\eqn\wh{
w^{(n)}_{zz} = \sqrt{ |n| (n^2 - 1 ) \over 2\pi} z^{|n|-2}~,
}
normalized such that 
\eqn\wi{\eqalign{
\int | w^{(n)}_{zz} |^2 \sqrt{g} d^2 z & = 1~.
}}
With this normalization the sum over all tensors give 
\eqn\wj{\eqalign{
\sum_{n=2}^\infty \bigg( |w^{(n)}_{zz} |^2 + {\rm c.c.}  \bigg)&= {1\over 2a^2} \sum_{n=-1}^\infty (1-|z|^2)^4 
\cdot { n(n^2 - 1 ) \over 2\pi} \cdot |z|^{2(n-2)}\cr
 & = {1\over 4\pi a^2 } (1-x)^4 \partial^3_x  {1\over 1 - x} =  {3\over 2\pi a^2}~. 
}}
This is three times the corresponding value for the normalized vector field derived from a non-normalizable scalar. In that case we referred to a single boundary mode so we interpret the result for the tensor as three boundary modes. There are of course infinitely many boundary modes enumerated by the index $n$ but there are three per unit volume.


\newsec{ Appendix C: Gravitino Modes on the Boundary.}

We want to find normalizable pure gauge gravitini constructed out of non normalizable spinor parameters. We start in analogy with the tensor boundary modes, studying the non normalizable solutions to Dirac's equation in AdS$_2$. 

We choose the same gamma matrices as Sen \SenBA\ for easy reference: 
\eqn\yab{\eqalign{
\gamma^{\hat{\theta}} &= - \sigma^2~, \cr
 \gamma^{\hat{\eta}} &= \sigma^1~. 
}}
We compute the twisted derivatives
\eqn\yabe{\eqalign{
D_\eta + {1\over 2}\gamma_\eta &= \partial_\eta + {1\over 2} \sigma^1~, \cr
D_\theta + {1\over 2}\gamma_\theta&= \partial_\theta + {i\over 2}\cosh\eta \sigma^3- {1\over 2}\sinh\eta \sigma^2~.
}}
The Dirac operator in the coordinates \bt\ with the gamma matrices \yab\ is
\eqn\yac{ \Dslash = -\sigma^2 {1\over \sinh \eta}\partial_\theta + \sigma^1 \partial_\eta +{1\over 2}\sigma^1\coth\eta
}
We will work with $a=1$ for now and restore it later. Camporesi and Higuchi \CamporesiFB, found the solutions
\eqn\yad{
\chi^{\pm}_k(\lambda) = e^{i(k+{1\over2})\theta}\pmatrix{ i{\lambda\over k+1} \cosh^{k}{\eta\over 2} \sinh^{k+1}{\eta\over 2}F(k+1+i\lambda, k+1-i\lambda;k+2; -\sinh^2{\eta\over 2}) \cr \pm \cosh^{k+1}{\eta\over 2} \sinh^{k}{\eta\over 2}F(k+1+i\lambda, k+1-i\lambda;k+1; -\sinh^2{\eta\over 2})}
}
and
\eqn\yae{
\eta^{\pm}_k(\lambda) = e^{-i(k+{1\over2})\theta}\pmatrix{\cosh^{k+1}{\eta\over 2} \sinh^{k}{\eta\over 2}F(k+1+i\lambda, k+1-i\lambda;k+1; -\sinh^2{\eta\over 2}) \cr \pm i{\lambda\over k+1} \cosh^{k}{\eta\over 2} \sinh^{k+1}{\eta\over 2}F(k+1+i\lambda, k+1-i\lambda;k+2; -\sinh^2{\eta\over 2})}
}
which satisfy
\eqn\yaea{\eqalign{
\Dslash ~\chi^{\pm}_k(\lambda) = \pm i \lambda \chi^{\pm}_k(\lambda)~, \cr
\Dslash ~\eta^{\pm}_k(\lambda) = \pm i \lambda \eta^{\pm}_k(\lambda)~.
}}
The label $k$ is a non-negative integer. The continuous spectrum is given by $\lambda$ real and positive. However, these are not all the modes of the Dirac operator, for there are non normalizable discrete modes with imaginary $\lambda$. The solution corresponding to $m^2 =1$ is $\lambda=i$. In this case the hypergeometric functions in \yad\ and \yae\ simplify,

\eqn\yaf{
\chi^{\pm}_k(i) = e^{i(k+{1\over2})\theta}\pmatrix{ - \sinh{\eta\over 2} \tanh^k{\eta\over 2} \cr \pm {1\over 2 \cosh{\eta\over 2} } (1+2k+\cosh\eta)\tanh^k {\eta\over 2} }~,
}

\eqn\yag{
\eta^{\pm}_k(i) = e^{-i(k+{1\over2})\theta}\pmatrix{   {1\over 2 \cosh{\eta\over 2} } (1+2k+\cosh\eta)\tanh^k {\eta\over 2}\cr \mp \sinh{\eta\over 2} \tanh^k{\eta\over 2} }~,
}

For $k \geq 0$. From now on we will refer to the solutions \yaf\ and \yag\ as $\chi^{\pm}_k$ and $\eta^{\pm}_k$ for simplicity, since we are interested in $m^2 =1$. 

Using the complex coordinates defined in \bt, the solutions \yaf\ and \yag\ are
\eqn\ybc{
\chi^\pm_k = \pmatrix{ -(1-|z|^2)^{-{1\over2}}|z|^{1\over2} \cr \pm  (1-|z|^2)^{{1\over2}}|z|^{-{1\over2}} (k+ {1 \over 1-|z|^2}) }z^{k + {1\over2}}
}
\eqn\ybd{
\eta^\pm_k = \pmatrix{ (1-|z|^2)^{{1\over2}}|z|^{-{1\over2}} (k+ {1 \over 1-|z|^2})  \cr \mp(1-|z|^2)^{-{1\over2}}|z|^{1\over2}}\bar{z}^{k + {1\over2}}
}
The normalization condition for the spinors \ybc\ and \ybd\ is
\eqn\ybda{
\int \bigg[{ |z| \over  1-|z|^2} +  {1-|z|^2 \over |z|}\bigg(k+ {1 \over 1-|z|^2}\bigg)^2 \bigg] |z|^{2k+1} {2 \over (1- |z|^2)^2 }  d^2z = \infty~.
}
These are non normalizable modes. We want to construct gravitini solutions that are pure gauge with gauge function proportional to the discrete modes \ybc\ and \ybd.

To construct the gravitini solutions we write the derivatives 
\eqn\ybg{\eqalign{
z D_z &  = z \partial_z + {1\over 4} {1+ |z|^2\over 1-|z|^2}\sigma^3 ~, \cr 
}}
and the holomorphic gamma matrix,
\eqn\ybh{ \eqalign{
z \gamma_z &= {|z| \over 1-|z|^2} ( \sigma^1 + i \sigma^2)~.
}}
Evaluation of the twisted holomorphic derivative yields
\eqn\ybk{
(D_z + {1\over 2} \gamma_z) \chi^+_k = \pmatrix{ 0 \cr 1}  k(k+1) \bigg({ 1-|z|^2 \over |z| }\bigg)^{{1\over2}} z^{k - {1\over2}}~.
}
 \ybk\ is explicitly convergent at $|z| \rightarrow 1$. Since the normalization integral for gravitini can be evaluated with the unit metric on the disk, we already know \ybk\ is normalizable. This is an advantage of working with complex coordinates. We compute the norm of \ybk,
\eqn\ybl{\eqalign{
\int  k^2(k+1)^2 \bigg({ 1-|z|^2 \over |z| }\bigg) |z|^{2k -1} d^2z &= 2\pi k^2(k+1)^2  \int_0^1   \bigg({ 1-x \over \sqrt{x} }\bigg) x^{k -{1\over2}} dx \cr
&= 2\pi k(k+1)~.
}}
The normalized gravitino boundary mode is
\eqn\ybm{
\Psi_z = \pmatrix{ 0 \cr 1} \sqrt{ { k(k+1) \over 2\pi } }\bigg({ 1-|z|^2 \over |z| }\bigg)^{{1\over2}} z^{k - {1\over2}}~.
}
The gravitini $\Psi_z$ are given for $k > 0$, since $k=0$ is explicitly zero. The solutions \ybm\ are normalizable modes that are pure gauge with a non normalizable gauge parameter. They are gravitino boundary modes. 

Through a similar computation one finds the modes $(D_z + {1\over 2} \gamma_z) \chi^-_k$ to be non normalizable. Also, if one computes the norms of $(D_z - {1\over 2} \gamma_z) \chi^\pm_k$ in analogy with the previous case, one finds that the gravitini $(D_z - {1\over 2} \gamma_z) \chi^+_k$ are non normalizable, while $(D_z - {1\over 2} \gamma_z)\chi^-_k$ are.

This is easily seen by noting that
\eqn\yar{
\chi_k^+ = \sigma^3 \chi^-_k~.
}
Also, according to \yab,
\eqn\yas{\eqalign{
[D_\mu, \sigma^3] &= 0~, \cr
\{\gamma_\mu, \sigma^3\} &= 0~.
}}
So that going from $(D_z + {1\over 2} \gamma_z)$ to $(D_z - {1\over 2} \gamma_z)$ can be achieved by multiplication with $\sigma^3$, which takes $\chi^+_k$ into $\chi^-_k$ and vice-versa. In fact, $(D_z - {1\over 2} \gamma_z) \chi^-_k$ is given by 
\eqn\yasa{
(D_z - {1\over 2} \gamma_z) \chi^-_k = \pmatrix{ 0 \cr - 1}  k(k+1) \bigg({ 1-|z|^2 \over |z| }\bigg)^{{1\over2}} z^{k - {1\over2}}~.
}
These are the modes \ybk\ up to a multiplicative constant. Thus, one should not count them as additional modes.

We find the action of the antiholomorphic twisted derivative on $\chi^+_k$ to vanish:
\eqn\yata{\eqalign{
&(D_{\bar{z}} + {1\over 2}\gamma_{\bar{z}}) \chi^+_k = 0~.
}}
When building gravitini out of the $\eta^\pm_k$ solutions, we find 
\eqn\yatb{\eqalign{
(D_{\bar{z}} + {1\over 2} \gamma_{\bar{z}}) \eta^+_k &= \pmatrix{ 1 \cr 0}  k(k+1) \bigg({ 1-|z|^2 \over |z| }\bigg)^{{1\over2}} \bar{z}^{k - {1\over2}}~, \cr
(D_z + {1\over 2} \gamma_z)  \eta^+_k&= 0~.
}}
and $(D_{\bar{z}} + {1\over 2} \gamma_{\bar{z}}) \eta^-_k $ are non normalizable. The normalized antiholomorphic modes are 
\eqn\yatc{
\bar{\Psi}_{\bar{z}} = \pmatrix{ 1 \cr 0} \sqrt{ { k(k+1) \over 2\pi } }\bigg({ 1-|z|^2 \over |z| }\bigg)^{{1\over2}} \bar{z}^{k - {1\over2}}~,
}
for $k>0$. The modes $\eta^-_k=\sigma^3\eta^+_k$ are once again just \yatb\ up to a phase.

In summary, the boundary modes we need to account for are \ybm\ and \yatc. One important property of these modes is that they are (anti-)holomorphic differentials:
\eqn\yatd{\eqalign{
&D_{\bar{z}}  \Psi_{z} =0~, \cr
&D_z \bar{\Psi}_{\bar{z}} = 0~.
}}
We have encountered a similar dependence for the tensor modes in \wh. The gravitini modes are different in that they are not powers of $z$ or $\bar{z}$, but instead have a $|z|$ dependent prefactor that is canceled by the spin connection.

 Finally, we sum over all values of $k$ in our boundary modes.
\eqn\ybn{\eqalign{
\sum_{k=1}^\infty \bigg( |\Psi_z|^2 + |\bar{\Psi}_{\bar{z}}|^2 \bigg) & = 2 \sum_{k=1}^\infty { k(k+1) \over 2\pi a^2} {(1-|z|^2)^3 \over 2} |z|^{2k -2}~, \cr
& = {1 \over 2\pi a^2 }  \sum_{k=-1}^\infty (1-x)^3 \partial^2 _x x^{k+1 }~,\cr
& = {2 \over 2\pi a^2 } ~.
}}
In the second equality we used the variable $x = |z|^2$, and added the empty entries $k=0,-1$. In the last step we evaluated the geometric series and the partial derivatives. We have one mode per unit volume for the holomorphic gravitino \ybm\ and one other mode for the antiholomorphic gravitino \yatc. 

The four boundary modes accounted for in Section 7 are the modes in \ybn\ times two supersymmetries.


\newsec{Appendix D: Conventions for Gamma-matrices.}
In this appendix we summarize our conventions, notations, and properties of gamma-matrices. 

The upper case $\Gamma_I$ refers to the 4D gamma matrices, while the lower case $\gamma^\mu$, $\gamma^\alpha$ refer to AdS$_2$ and $S^2$, respectively. They satisfy:
\eqn\xa{\eqalign{
 & ~~~~~~\{\Gamma^I, \Gamma^J\} =2g^{IJ}~, \cr
\Gamma^\mu &= \gamma^\mu\otimes \gamma_S ~, ~~~~
\Gamma^\alpha =  1 \otimes \gamma^\alpha~, \cr
& ~~~~~~ [\gamma^\mu,\gamma^\alpha]=0~,
}}
Chiral projection operators in 4D and 2D, along with their relations:
\eqn\xb{\eqalign{
\Gamma_5   &= i\Gamma^0 \Gamma^1 \Gamma^2 \Gamma^3 = \gamma_A \otimes \gamma_S~,\cr
\gamma_A &= \gamma^{01} , ~~~~\gamma_S =  i\gamma^{23}~,\cr
& [\gamma_A,\gamma_S]=0~, ~~~~\cr
& \gamma_A^2 = \gamma_S^2 = 1~.
}}
Conventions on orientation (all indices are local)
\eqn\xc{\eqalign{
& ~\epsilon^{0123}  = +1~,\cr
\epsilon^{01}  =& +1~, ~~~~ \epsilon^{23}  = +1~.
}}
Some useful identities,
\eqn\xd{\eqalign{
\Gamma^{IJKL}  = - i \Gamma_5 \epsilon^{IJKL}&, ~~~~\Gamma^{IJK} = - i \Gamma_5 \epsilon^{IJKL}\Gamma_L~, \cr
\gamma_A \epsilon^{\mu\nu}  = \gamma^{\mu\nu}& , ~~~~ \gamma_S \epsilon^{\alpha\beta}  = i\gamma^{\alpha\beta}~.
}}

\listrefs

\end